\documentclass[onecolumn,preprintnumbers,pre]{revtex4}
\usepackage{graphicx, subfigure}
\usepackage{amssymb, amsmath,amssymb,amsfonts}
\usepackage{amsthm,mathrsfs,amsopn}
\usepackage{dcolumn}
\usepackage{bm}
\usepackage{color}
\usepackage[utf8]{inputenc}
\usepackage{mathtools}

\theoremstyle{plain} 

\begin{document}

\title{Non-reciprocal interactions {enhance} {heterogeneity}}

\author{Timoteo Carletti}
\email{timoteo.carletti@unamur.be}
\affiliation{Department of Mathematics and naXys, Namur Institute for Complex Systems, University of Namur, Rue Graf\'e 2, B5000 Namur, Belgium}

\author{Riccardo Muolo}
\affiliation{Department of Mathematics and naXys, Namur Institute for Complex Systems, University of Namur, Rue Graf\'e 2, B5000 Namur, Belgium}


\begin{abstract}
We study a process of pattern formation for a {generic} model of species anchored to the nodes of a network where local reactions take place, and that experience non-reciprocal non-local long-range interactions{, encoded by the network directed links}. {By assuming the system to exhibit} a stable homogeneous equilibrium whenever only local interactions are considered, we prove that such equilibrium can turn unstable once suitable non-reciprocal non-local long-range interactions are allowed for. Stated differently we propose sufficient conditions allowing for patterns to emerge by using a non-symmetric coupling, while initial perturbations about the homogenous equilibrium always fade away by assuming reciprocal coupling, namely the latter is stable. The instability, precursor of the emerging  spatio-temporal patterns, can be traced back, via a linear stability analysis, to the complex spectrum of an interaction non-symmetric Laplace operator. The proposed theory is then applied to several paradigmatic dynamical models largely used in the literature to study the emergence of patterns or synchronisation. Taken together, our results pave the way for the understanding of the {many and  {heterogeneous} patterns of complexity} found in ecological, chemical or physical systems composed by interacting parts, once no diffusion takes place.
\end{abstract}
\maketitle

\section{Introduction}
\label{sec:intro}
\noindent We live in an interconnected world~\cite{newmanbook,barabasibook} where complex patterns~\cite{pastorsatorrasvespignani2010} spontaneously emerge from the intricate web of nonlinear interactions existing between the basic units by which the system under study is made of~\cite{Nicolis1977}. These emergent structures can be found in the synchronised activity of neurones, resulting from the exchange of electrochemical signals via synapses~\cite{Pikovsky2001,arenasreview}, as well as the geometric visual hallucinations product of the retinocortical map linking the retina and the striate cortex~\cite{BCGTW,BC2003}. Such self-organised structures can also manifest in groups of fireflies that flash at unison, each one by observing the behaviour of their close neighbours~\cite{Strogatzbooksync}. Or they can materialise as the striped or spotted motifs on the skin of the zebrafish {\em Danio rerio}, due to the long-range interactions between melanophores and xantophores, without requiring diffusion nor any kind of cell motion~\cite{kondo2009,BullaraDeDecker}. 

The latter phenomena, as well as many other ones modelled within a similar framework of local reactions and long-range interactions without displacement  {of the reacting species}, cannot be ascribed to a Turing instability~\cite{Turing}, a widely used paradigm for pattern formation. Indeed the latter requires a diffusive process, whereas the common key factor linking the above examples is the immobility of the reacting species and the existence of a web of long-range interactions due to some signal propagation. 

In this work we set the theoretical basis for the understanding of self-organisation in systems  {without diffusing species} where asymmetric long-range interactions play a pivotal role. Network science~\cite{newmanbook,barabasibook} provides a natural framework where to study such phenomena. Indeed local interactions can be described by a dynamical system evolving on each node, that represents a portion of physical space, say a natural habitat or a chemical reactor, large enough to contain enough species to describe their evolution with a nonlinear ordinary differential equation. On the other hand, non-local long-range interactions can be modelled by links connecting the nodes of the network allowing thus species anchored to each distinct node to communicate by exchanging some generic signals. Let us observe that our definition of local and long-range dynamics can differ from the one sometimes used in the literature, that considers local dynamics to be related to the in-node processes but also to the interactions with species sitting in nodes at distance one, i.e., first neighbourhood, while long-range interactions are used to describe far away interacting nodes, i.e., distances larger than two. For this reason in this work we name the latter process non-local long-range interaction; let us notice however that when it will be clear from the context, we will simply use long-range interactions. Let us finally stress that the latter do not involve displacement of the reacting species, for this reason the theory hereby developed applies to systems without diffusion. An interesting example is the web of chemical light-triggered reactions obtained by connecting reactors where the local concentration of chemicals, i.e., in each node, determines the amount of light to be put onto connected distant nodes, i.e., long-range interaction. Chemicals do not leave any nodes but they interact at distance~\cite{GizynskiGorecki2017,C7CP03260A,PGST2020}.

In the aforementioned examples, the system converges to an homogeneous solution, being stationary or time varying, once only local interactions are taken into account, i.e., long-range ones are silenced, while heterogeneous solutions spontaneously emerge in presence of a suitable web of interactions among the units.  {Let us observe that the existence of an attracting homogeneous solution for the decoupled system is a natural and largely assumed working hypothesis, see for instance~\cite{Turing} in the framework of Turing instability or~\cite{Pecora} for synchronisation.}

A preliminary result in this direction has been proposed in~\cite{CENCETTI2020109707} with the assumption of reciprocal non-local interactions and by treating the latter in a mean-field setting. There, authors have been able to prove that the stable homogeneous solution, existing once the long-range interactions have been silenced, can turn unstable by introducing a suitable  {symmetric} non-local coupling and eventually lead to the emergence of a spatially (and temporally) dependent solution. The aim of the present work is to make one step forward by studying the general case of {\em non-reciprocal} interactions  {and their impact on the system outcome}. Indeed, the interactions existing among the constituting units are often not symmetric; this is the case of plant-animal mutualistic networks~\cite{Bascompte,AbramsonTrejoSotoOna},  {the specific example of sheeps and deers~\cite{EPBMCW}} or the case of olfactory receptor neurones in the Drosophila antenna~\cite{YeSu2019}, just to mention a few.

Anticipating our conclusion, we claim that the diversity  {and heterogeneity of patterns} observed in Nature, being associated to spatial or temporal  {non-homogeneous states}, is {enhanced}  by non-reciprocal long-range interactions. The onset of the instability, precursor of the pattern, can be detected with a linear stability analysis, providing a condition on the complex spectrum of a non-symmetric consensus Laplacian operator {resulting from the mean-field ansatz as we will hereafter explain}. The proposed framework is general enough to cover systems {of any dimension $d\geq 2$} displaying a fixed-point or a limit cycle homogeneous solution, once we silence the long-range interactions, by proving the existence of a suitable non-reciprocal web of interactions driving the emergence of patterns while the latter can never exist in the case of symmetric support. To emphasise the relevance of the proposed theory to many and different research domains, we applied the developed theory to several paradigmatic models largely used in the literature to study the emergence of patterns or synchronisation. In conclusion, the proposed mechanism provides the way for alternative routes to pattern formation, beyond the Turing one~\cite{Turing,NM2010},  {and to the emergence of desynchronised states~\cite{Pecora}}, suitable for all phenomena where diffusion is not the main driver for the heterogeneity of the complex patterns seen in Nature, opening thus new possibilities for modelling ecological, chemical and physical interacting systems, endowed with non-reciprocal couplings {and without diffusing species}.

\section{The model}
\label{sec:model}
\begin{figure}[t]
\includegraphics[scale=.2]{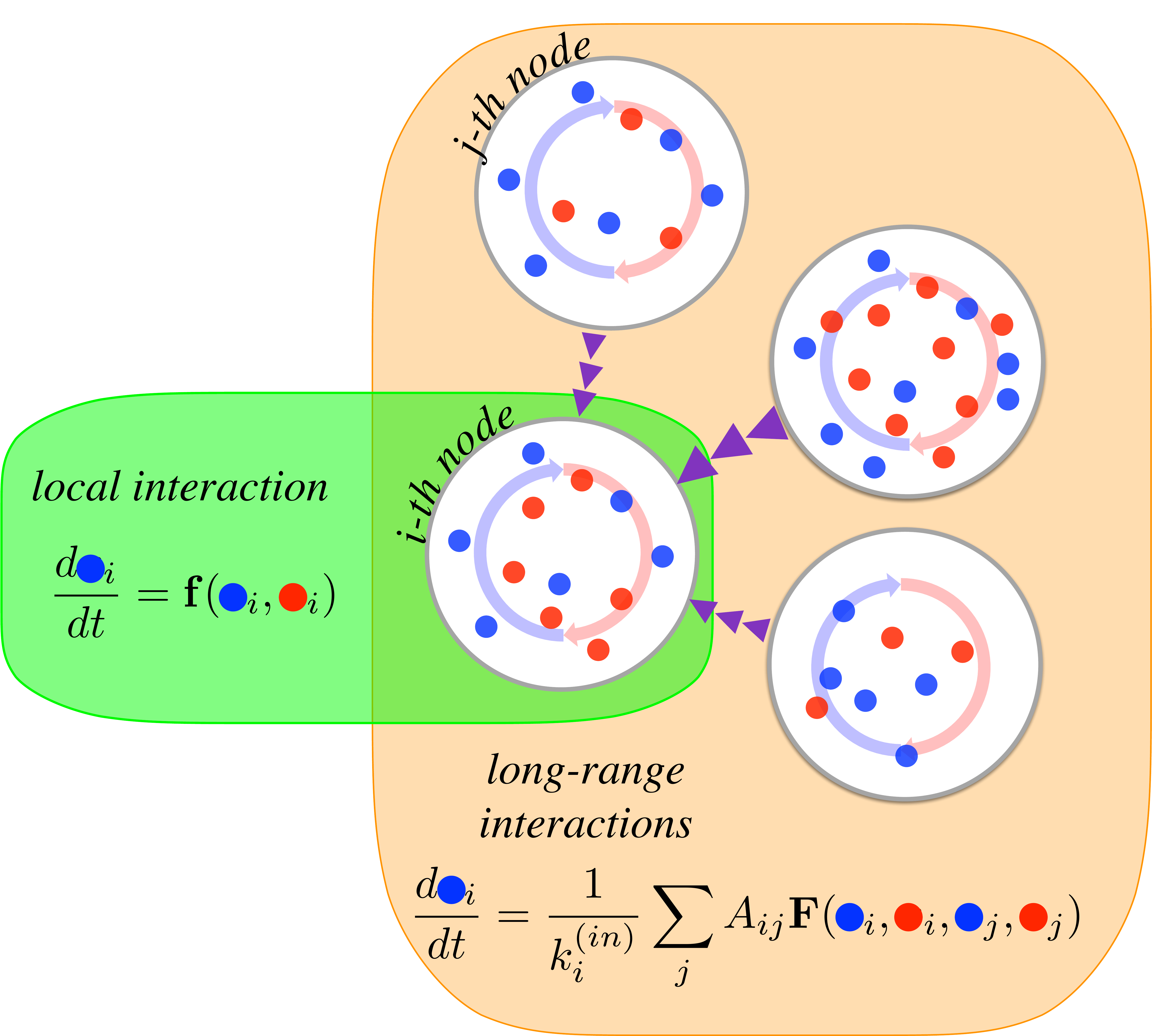}
\caption{ {\textbf{A schematic visual representation of the model~\eqref{eq:sysnA}.} Nodes (large white circles) contain two species (blue and red dots). We then focus on the growth rate of the ``blue'' species in the $i$-th node, that results from two terms; the first one is the local reaction $\mathbf{f}$, represented by the light-blue and light-red curved arrows, involving only ``blue'' and ``red'' species indexed by $i$ (green oval). The second contribution arises from the long-range interactions $\mathbf{F}$, represented as violet triangles pointing from each nearby node to the $i$-th one. The average of these terms impacts the growth rate of the ``blue'' specie in node $i$ (orange oval).}}
\label{fig:themodel} 
\end{figure}

Let us consider a dynamical system composed by $n$ identical parts and assume the $d$-dimensional vector, $\mathbf{x}^{(i)}(t)=(x^{(i)}_1(t),\dots,x^{(i)}_d(t))^\top$, to represent the state of the $i$-th copy,  {for $i=1,\dots,n$}. The isolated systems are described by an  {ordinary differential equation~\footnote{\textcolor{black}{Let us observe that the theory hereby developed could be straightforwardly adapted to the case of discrete time, where, i.e., the ODE is replaced with a finite differences system.}}, resulting from the assumption of a ``well stirred'' distribution of species inside each node}
\begin{equation}
\label{eq:sysn}
 \frac{d\mathbf{x}^{(i)}}{dt}=\mathbf{f}(\mathbf{x}^{(i)})\quad\forall i=1,\dots, n\, ,
\end{equation}
where $\mathbf{f}$ is a generic nonlinear function responsible for the local interactions,  {i.e., depending on the species anchored to the same node}. Let us now allow each system to possess non-local interactions, i.e.,  {the growth rate of the system anchored at node $i$ is influenced by some nonlinear function of the amount of species in distant nodes}. Moreover we assume the latter to be described by
\begin{equation}
 \label{eq:sysnA}
 \frac{d\mathbf{x}^{(i)}}{dt}=\frac{1}{k_i^{(in)}}\sum_j A_{ij} \mathbf{F}(\mathbf{x}^{(i)},\mathbf{x}^{(j)})\quad\forall i=1,\dots, n\, ,
\end{equation}
where $A_{ij}$ is the (possibly weighted) non-symmetric adjacency matrix encoding the long-range interactions, i.e., $A_{ij}=1$ if and only if node $j$ influences node $i$. Let $k_i^{(in)}=\sum_j A_{ij}$ be the in-degree of node $i$ and observe that we allow $A_{ii}=1$, thus the in-degree takes into account also the possible self-loops. Finally, $\mathbf{F}(\mathbf{x}^{(i)},\mathbf{x}^{(j)})$ is a nonlinear function that describes the effect of the $j$--th system on the $i$-th one. Moreover we require $\mathbf{F}(\mathbf{x}^{(i)},\mathbf{x}^{(i)})=\mathbf{f}(\mathbf{x}^{(i)})$, namely the self-interaction is represented by the original nonlinear function $\mathbf{f}$ describing the evolution of the isolated systems. This is a natural assumption allowing to recover the initial system~\eqref{eq:sysn} composed by isolated units, i.e., once we set $\mathbf{A}=\mathbf{I}_n$, being the latter the $n\times n$ identify matrix~\footnote{Let us observe that this condition has been already used in~\cite{CENCETTI2020109707} and it is related to the one adopted in~\cite{GdPGLRCFLB2020}.}. Let us observe that the right hand side of~\eqref{eq:sysnA} can be rewritten as the average of the interactions perceived by node $i$, $\langle \mathbf{F}(\mathbf{x}^{(i)},\cdot)\rangle= \sum_j A_{ij} \mathbf{F}(\mathbf{x}^{(i)},\mathbf{x}^{(j)})/k_i^{(in)}$, hence describing the mean-field ansatz.  {Finally, let us stress that species cannot move across nodes: the long-range interactions are thus mediated by some generic signals, as we have proposed in the introduction and we will discuss in the following. The model~\eqref{eq:sysnA} describes thus local interactions coupled with long-range ones without the need of diffusion, this is in contrast with the fundamental assumption of Turing instability~\cite{Turing,NM2010} where diffusion plays a key role, or in the synchronisation based on diffusive-like couplings~\cite{Pecora,Pecora_etal97} (see Fig.~\ref{fig:themodel} for a schematic representation of the proposed model).}

By defining the matrix 
\begin{equation}
\label{eq:consLapHG}
\mathcal{L}_{ij}=\frac{A_{ij}}{k^{(in)}_i}-\delta_{ij}\, ,
\end{equation}
and by using the property of the function $\mathbf{F}$, we can rewrite Eq.~\eqref{eq:sysnA} as
\begin{equation}
 \label{eq:sysnAL}
 \frac{d\mathbf{x}^{(i)}}{dt}=\mathbf{f}(\mathbf{x}^{(i)})+\sum_j \mathcal{L}_{ij} \mathbf{F}(\mathbf{x}^{(i)},\mathbf{x}^{(j)})\quad\forall i=1,\dots, n\, .
\end{equation}
Once the long-range interactions are silenced and each node interacts only with itself, then $\mathcal{L}=0$ and thus Eq.~\eqref{eq:sysnAL} reduces to~\eqref{eq:sysn}. Let us observe that in the case of reciprocal interactions, the above defined matrix $\mathcal{L}$ corresponds to the {\em consensus Laplace} operator~\cite{Ghosh_etal14,Lambiotte_etal11,OlfatiSaber07,Krause09}, also named {\em reactive Laplace} matrix in~\cite{CENCETTI2020109707}, whose spectrum is real and non-positive. In the case under scrutiny, involving non-reciprocal interactions, the spectrum is generally complex but one can prove that $\Lambda^{(1)}=0$ is still an eigenvalue associated to the uniform eigenvector $\phi^{(1)}\sim (1,\dots,1)^{\top}$. Moreover the Gershgorin circle theorem~\cite{Ger31} allows to prove that the real part of the spectrum of $\mathcal{L}$ is contained in the strip $[-2,0]$ in the complex plane, hence $\mathcal{L}$ is stable having all the eigenvalues but $0$ with a negative real part. Observe that by using a symmetric network the eigenvalues are real and non-positive, even if the consensus Laplace is not symmetric (see {Appendix}~\ref{sec:spectLapsymm}). {Let us stress that the operator $\mathcal{L}$ results from the mean-field assumption and not from any kind of diffusive-like process as in the cases where the combinatorial Laplace operator arises because of the Fick law.}

Let us assume $\mathbf{x}^{(i)}(t)=\mathbf{s}(t)$, $i=1,\dots,n$, to be a solution of the initial system~\eqref{eq:sysn}; then because of the above hypothesis on $\mathbf{F}$ and of the definition of $k_i^{(in)}$, it is also a spatially{, i.e., node} dependent solution of Eq.~\eqref{eq:sysnAL}. To study the bifurcation of patchy solutions from the stable homogeneous one,  {$\mathbf{x}^{(i)}(t)=\mathbf{s}(t)$ for all $i$}, we consider a node dependent perturbation about the latter  {solution},  {i.e.,} $\mathbf{u}^{(i)}(t)=\mathbf{x}^{(i)}(t)-\mathbf{s}(t)$, whose evolution can be studied by inserting it into Eq.~\eqref{eq:sysnAL} and then keeping only first order terms, assumed to be small enough. Hence  {we obtain} for all $i=1,\dots, n$ 
\begin{equation}
\label{eq:linear}
 \frac{d\mathbf{u}^{(i)}}{dt}=\left[\mathbf{J}_1(\mathbf{s}(t))+\mathbf{J}_2(\mathbf{s}(t))\right]\mathbf{u}^{(i)}+\sum_j \mathcal{L}_{ij} \mathbf{J}_2(\mathbf{s}(t))\mathbf{u}^{(j)} \, ,
\end{equation}
where we have introduced the Jacobian matrices  {$\mathbf{J}_1(\mathbf{s})=\partial_{\mathbf{x}_1} \mathbf{F}(\mathbf{x}_1,\mathbf{x}_2)\rvert_{(\mathbf{s},\mathbf{s})}$}, i.e., the derivatives are computed with respect to the first group of variables, and  {$\mathbf{J}_2(\mathbf{s})=\partial_{\mathbf{x}_2} \mathbf{F}(\mathbf{x}_1,\mathbf{x}_2)\rvert_{(\mathbf{s},\mathbf{s})}$}, i.e., the derivatives are performed with respect to the second group of variables. In both cases the derivatives are evaluated on the reference solution $\mathbf{s}(t)$. 

This latter equation encodes $n$ linear systems involving matrices with size $d\times d$. To progress with the analytical understanding, we assume the existence of an eigenbasis~\footnote{{Let us observe that this hypothesis can be relaxed (see Appendix~\ref{sec:dynb}) and one could obtain similar results by invoking the Jordan block decomposition as done in~\cite{NM2006} in the case of synchronisation. For sake of definitiveness we preferred to present our results under this restrictive assumption, but allowing for straightforward analysis.}} of right eigenvectors $\phi^{(\alpha)}$, $\alpha=1,\dots, n$ for $\mathcal{L}$ and then, {following the ideas pioneered by~\cite{Turing,Pecora}}, we project the former equation onto each eigendirection, i.e., $\mathbf{u}^{(i)}=\sum_\alpha \mathbf{u}^{(\alpha)}\phi^{(\alpha)}_i$, to eventually obtain (see  {Appendix}~\ref{sec:dynb})
\begin{equation}
\label{eq:finalEqalpha}
 \frac{d\mathbf{u}^{(\alpha)}}{dt}=\left[\mathbf{J}(\mathbf{s}(t))+\mathbf{J}_2 (\mathbf{s}(t))\Lambda^{(\alpha)}\right]\mathbf{u}^{(\alpha)} \quad \forall \alpha=1,\dots, n\, ,
\end{equation}
where $\mathbf{J}=\mathbf{J}_1+\mathbf{J}_2$ and $\Lambda^{(\alpha)}$ is the eigenvalue relative to the eigenvector $\phi^{(\alpha)}$. The above equation enables us to infer the (in)stability of the homogeneous solution, $\mathbf{s}(t)$, by studying the Master Stability Function~\cite{Pecora,HCLP}, namely the largest Lyapunov exponent of Eq.~\eqref{eq:finalEqalpha}, \textcolor{black}{considered as a function of $\Lambda^{(\alpha)}$}. 

To make one step further in the study of the problem, let us hypothesise that each isolated system converges to the same stationary point, i.e., the stable homogeneous solution is  {time independent}, $\mathbf{s}(t)=\mathbf{s}_0$. Hence Eq.~\eqref{eq:finalEqalpha} rewrites for all $\alpha=1,\dots, n$, as
\begin{equation}
\label{eq:finalEqalphaHom}
 \frac{d\mathbf{u}^{(\alpha)}}{dt}=\left[\mathbf{J}(\mathbf{s}_0)+\mathbf{J}_2 (\mathbf{s}_0)\Lambda^{(\alpha)}\right]\mathbf{u}^{(\alpha)}:=\mathbf{J}^{(\alpha)} \mathbf{u}^{(\alpha)}\, ,
\end{equation}
 {Where the constant matrix $\mathbf{J}^{(\alpha)}$ has been defined by the latter equation.} The homogeneous solution will prove unstable to spatial dependent perturbations if (at least) one eigenmode $\hat{\alpha}$ exists for which the largest real part of the eigenvalues, $\lambda_i$, of $\mathbf{J}^{(\hat{\alpha})}$ is positive, the latter being known in the literature with the name of {\em dispersion relation}, hereby denoted by $\rho_\alpha=\max_{i=1,\dots,d}\Re\lambda_i(\Lambda^{(\alpha)})$, where we emphasised the dependence of the latter on the spectrum of the Laplace matrix.  {Let us observe that the positivity of the dispersion relation initiates the instability but it does not determine the final outcome of the system. Convergence to other homogeneous solutions is possible if $\mathbf{f}$ possesses several zeros, namely if the system~\eqref{eq:sysn} exhibits multiple equilibria. However, this kind of solution can arise with lower probability with respect to a patchy one; indeed, assuming to act on a parameter in Eq.~\eqref{eq:sysnAL}, then a homogeneous zero of the latter generically bifurcates in an heterogenous one.}

\section{Results}
\label{sec:res}
For sake of  {pedagogy and to be able to determine closed and manageable analytical formulas}, let us assume the local systems to be $2$ dimensional, i.e., $d=2$ in Eq.~\eqref{eq:sysn}. {Let us however emphasise that, as we will discuss later on, our results go beyond this simplified framework and we can indeed  prove that non-reciprocal interactions can drive the emergence of heterogenous patterns in any dimension $d\geq 2$, even when this is not possible using a symmetric coupling.}  {Assuming thus $d=2$,} being the eigenvalues of  $\mathbf{J}^{(\alpha)}$ the solutions of the second order equation
\begin{equation}
\label{eq:disprel2}
 \lambda_i^2-\mathrm{tr}\mathbf{J}^{(\alpha)}\lambda_i+\det \mathbf{J}^{(\alpha)}=0\, ,
\end{equation}
we can adapt to the present case the analysis done in~\cite{Asllani2014NC,ACFM2020} and express the condition for the onset of instability, i.e., $\rho_{\hat{\alpha}}>0$, as follows
\begin{equation}
\label{eq:instabcond}
\exists \hat{\alpha}>1\text{ st }
(\Im\Lambda^{(\hat{\alpha})})^2S_2(\Re\Lambda^{(\hat{\alpha})})\leq -S_1(\Re\Lambda^{(\hat{\alpha})})\, ,
\end{equation}
where $S_2(\xi)$, resp. $S_1(\xi)$, is a second, resp. fourth, degree polynomial in $\xi$ (see   {Appendix}~\ref{sec:dynb} for more details  {and the explicit form of $S_1$ and $S_2$ in terms of the model parameters}).

To illustrate the potential of the theory let us consider several examples of dynamical systems largely used in the literature as paradigmatic models for synchronisation or patterns emergence.

\subsection{The Brusselator model}
\label{ssec:Bxl}

Let us consider the Brusselator model~{\cite{PrigogineNicolis1967,PrigogineLefever1968}}, often invoked in the literature as {a paradigm} nonlinear reaction scheme for studying  self-organised phenomena, synchronisation~\cite{entropy}, Turing patterns~\cite{NM2010,Asllani2014,Asllani2016,PABFC2017,ACFM2020} and oscillation death~\cite{KVK2013,LFCP2018}. The key feature of the model is the presence of two species, reacting via a cubic nonlinearity
\begin{equation}
\label{eq:bxl}
\begin{dcases}
\frac{d u}{dt} &= 1-(b+1)u+c u^2v\\
\frac{d v}{dt} &= bu-c u^2v\, ,
\end{dcases}
\end{equation}
where $b>0$ and $c>0$ act as tunable model parameters. One can easily realise the existence of a unique equilibrium $u^* = 1$ and $v^* = b/c$, that results stable if the Jacobian of the reaction part evaluated on the equilibrium $\mathbf{J}_{\mathrm{Bxl}}=\left( 
\begin{smallmatrix}
b-1 & c \\-b & -c 
\end{smallmatrix}\right)$, has a negative trace, $\mathrm{tr}\mathbf{J}_{\mathrm{Bxl}}=b-c<1$, and a positive determinant $\det \mathbf{J}_{\mathrm{Bxl}}=c>0$.

The model can be cast in the framework presented in the previous section by setting $\mathbf{x}^{(i)}=(u_i,v_i)$ and $\mathbf{f}(\mathbf{x}^{(i)})=(1-(b+1)u_i+c u_i^2v_i,bu_i-c u_i^2v_i)$, for $i=1,\dots,n$ to denote the $n$ isolated systems. For sake of concreteness let us consider the coupling given by
\begin{equation*}
 \mathbf{F}(\mathbf{x}^{(i)},\mathbf{x}^{(j)})=(1-(b+1)u_i+c u_i^2v_i,bu_j-c u_i^2v_i)\, ,
\end{equation*}
where $\mathbf{x}^{(j)}=(u_j,v_j)$. Such function clearly satisfies the constraint $ \mathbf{F}(\mathbf{x}^{(i)},\mathbf{x}^{(i)})=\mathbf{f}(\mathbf{x}^{(i)})$, hence Eq.~\eqref{eq:sysnA} becomes (see Appendix~\ref{sec:bxlmod})
\begin{equation}
 \label{eq:sysnABxl}
 \begin{dcases}
\frac{d u_i}{dt} &= 1-(b+1)u_i+c u_i^2v_i\\
\frac{d v_i}{dt} &= bu_i-c u_i^2v_i+b\sum_j \mathcal{L}_{ij}u_j
\end{dcases} \quad\forall i=1,\dots, n\, ,
\end{equation}
namely a set of $n$ Brusselator models~\eqref{eq:bxl} coupled via long-range connections in the second variable.

\begin{figure}[t]
\includegraphics[scale=.3]{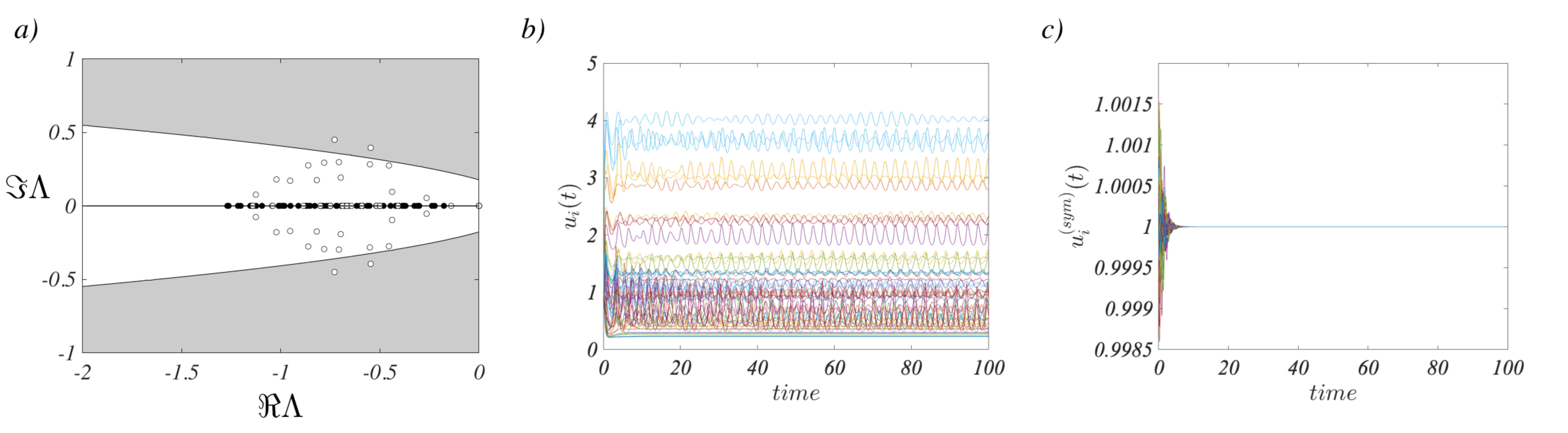}
\caption{\textbf{Instability region and patterns for the Brusselator model}. In panel a) we report the region of the complex plane $(\Re \Lambda,\Im \Lambda)$ for which the instability condition is satisfied (grey). For the chosen parameters values ($b = 4.3$ and $c= 5.0$) we can observe that the instability region does not intersect the real axis and thus only non-reciprocal coupling can exhibit complex eigenvalues (white dots) entering into the instability region and thus initiate the pattern as shown in panel b) where we report $u_i(t)$ vs $t$ starting from initial conditions close to the stable equilibrium, $u_*=1$. Any symmetric coupling determines real eigenvalue (black dots in panel a)) that cannot give rise to the instability, as shown in panel c) where we report $u_i(t)$ vs $t$ starting from the same initial conditions used in panel b). The underlying coupling is obtained with a directed Erd\H{o}s-R\'enyi network with $n=50$ nodes and a probability for a direct link to exist between two nodes is $p=0.05$.}
\label{fig:Bxl}
\end{figure}

Given the above coupling we can explicitly compute the polynomials $S_1(\xi)$ and $S_2(\xi)$ as a function of the model parameters (see {Appendix}~\ref{sec:bxlmod}) and characterise the instability region defined by~\eqref{eq:instabcond} as reported in panel a) of Fig.~\ref{fig:Bxl} where we show in the complex plane $(\Re \Lambda,\Im \Lambda)$ the regions for which the instability condition is satisfied (grey), for a given set of parameters. Patterns do emerge if there exists at least one eigenvalue $\Lambda^{(\hat{\alpha})}$ belonging to this region. For sake of simplicity, we hereby assume the non-reciprocal interactions to be described by a directed Erd\H{o}s-R\'enyi network made of $50$ nodes and the probability to create a directed link to be $0.05$. The symmetric coupling is obtained by considering all the existing links to be reciprocal ones and, as already noticed, the eigenvalues are negative real numbers. In conclusion, if the model parameters shape an instability region that does not intersect the real negative axis (see panel a) in Fig.~\ref{fig:Bxl}), then only an asymmetric coupling can drive the instability and the ensuing (oscillatory) patterns (see panel b) in Fig.~\ref{fig:Bxl}), while this is impossible for {any} web of reciprocal long-range interactions (see panel c) in Fig.~\ref{fig:Bxl}) and the system solution converges toward the homogenous solution~\footnote{Throughout the work, the numerical simulations have been performed by initialising the system $\delta$-close to the homogeneous equilibrium by drawing uniformly random perturbations in $(-\delta,\delta)$. Then its time evolution has been numerically simulated using a $4$-th order Runge-Kutta method over a time span of the order of $-\log \delta /\max_\alpha \rho_\alpha$, namely sufficiently long to (possibly) increase the $\delta$-perturbation up to a macroscopic size.}. 

\subsection{The Mimura-Murray model}
\label{ssec:MM}

The second model we consider is the Mimura-Murray system~\cite{MimuraMurray1978,NM2010,chall,ABCFP2015}. It also involves two reacting species, that can be associated to prey and predator, whose densities denoted by $u$ and $v$ evolve via the nonlinear system of ODEs:
\begin{equation}
\label{eq:fgMM}
\begin{dcases}
 \frac{du}{dt} &=u\left( \frac{a+bu-u^2}{c} - v\right)\\
  \frac{dv}{dt} &=v\left( u - (1+dv) \right)\, ,
\end{dcases}
\end{equation}
where $a, b, c$ and $d$ are positive parameters. The model possesses $6$ equilibria, whose stability and positivity depend on the value of the chosen parameters. We here focus on the fixed point $(u_*,v_*)$
\begin{equation}
\label{eq:eqMM}
u_*= 1+\frac{bd - 2d - c + \sqrt{\Delta}}{2d}\text{ and }v_*= \frac{bd - 2d - c + \sqrt{\Delta}}{2d^2}\text{ where }\Delta = (bd-2d-c)^2 + 4d^2(a + b -1)\, ,
\end{equation}
and assume $a=35$, $b=15$, $c=20$ and $d=0.4$ which in turn implies $(u_*,v_*)\sim (2.28,3.20)$. The Jacobian matrix evaluated at the fixed point reads $\mathbf{J}_{\mathrm{MM}}\sim\left(
\begin{smallmatrix}
1.19 & -2.28\\3.20 &-1.28 
\end{smallmatrix}\right)$, hence, $\det \mathbf{J}_{\mathrm{MM}}>0$ and $\mathrm{tr}\mathbf{J}_{\mathrm{MM}}<0$ and the fixed point is a stable equilibrium. 

Let us consider the following long-range coupling
\begin{equation*}
 \mathbf{F}(\mathbf{x}^{(i)},\mathbf{x}^{(j)})=\left(u_i\left( \frac{a+bu_j-u_i^2}{c} - v_j\right),v_i\left( u_j - (1+dv_i) \right) \right)\, ,
\end{equation*}
where $\mathbf{x}^{(i)}=(u_i,v_i)$ and $\mathbf{x}^{(j)}=(u_j,v_j)$. Let us observe that the present coupling contains nonlinear terms in the variables $u_iv_j$ and $u_jv_i$, while in the previous section we dealt with a linear term. Let $\mathbf{f}(\mathbf{x}^{(i)}) = \left(u_i\left( \frac{a+bu_i-u_i^2}{c} - v_i\right),v_i\left( u_i - (1+dv_i) \right)\right)$, then $ \mathbf{F}(\mathbf{x}^{(i)},\mathbf{x}^{(i)})=\mathbf{f}(\mathbf{x}^{(i)})$, and we are thus in the framework of Eq.~\eqref{eq:sysnA} that now can be rewritten as (see Appendix~\ref{sec:mmmod})
\begin{equation}
 \label{eq:sysnAMM}
 \begin{dcases}
\frac{d u_i}{dt} &= u_i\left( \frac{a+bu_i-u_i^2}{c} - v_i\right)+\frac{b}{c}u_i\sum_j \mathcal{L}_{ij}u_j-u_i\sum_j \mathcal{L}_{ij}v_j\\
\frac{d v_i}{dt} &= v_i\left( u_i - (1+dv_i) \right)+v_i\sum_j \mathcal{L}_{ij}u_j
\end{dcases} \quad\forall i=1,\dots, n\, .
\end{equation}
One can realise that we are dealing with $n$ Mimura-Murray models~\eqref{eq:fgMM} coupled via nonlinear long-range connections in both variables.

\begin{figure}[t]
\includegraphics[scale=.3]{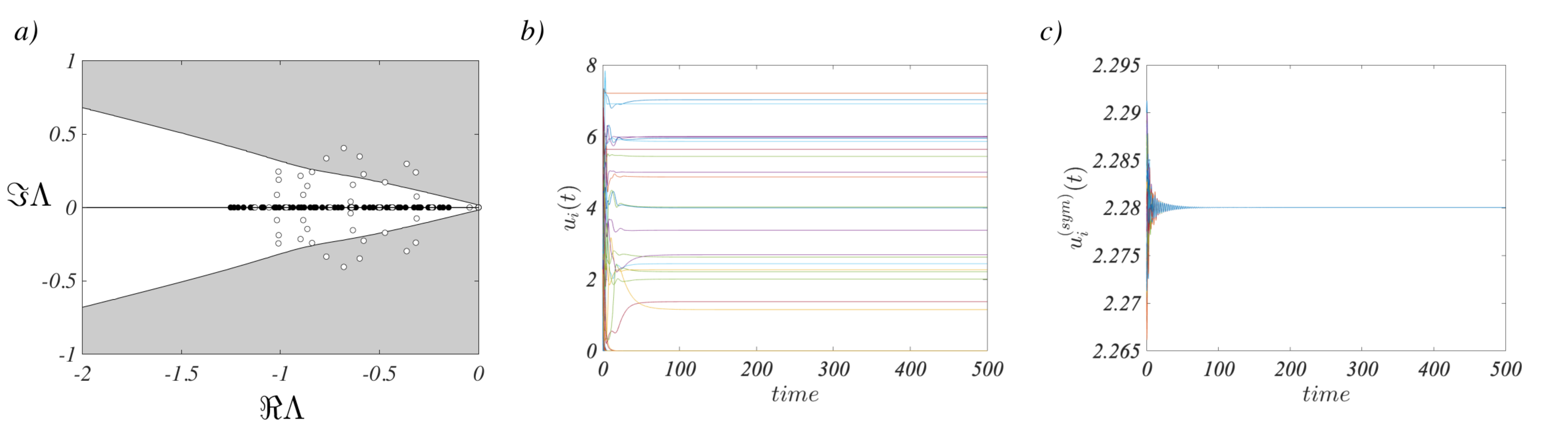}
\caption{\textbf{Instability region and patterns for the Mimura-Murray model}. In panel a) we report the region of the complex plane $(\Re \Lambda,\Im \Lambda)$ for which the instability condition is satisfied (grey). For the chosen parameters values ($a = 35$, $b=15$, $c= 20$ and $d=2/5$) we can observe that the instability region does not intersect the real axis and thus only non-reciprocal coupling can exhibit complex eigenvalues (white dots) entering into the instability region and thus initiate the pattern as shown in panel b) where we report $u_i(t)$ vs $t$ starting from initial conditions close to the stable equilibrium, $u_* \sim 2.28$, $v_*\sim 3.20$. Any symmetric coupling determines real eigenvalue (black dots in panel a)) that cannot give rise to the instability, as shown in panel c) where we report $u_i(t)$ vs $t$ starting from the same initial conditions used in panel b). The underlying coupling is obtained with a directed Erd\H{o}s-R\'enyi network with $n=50$ nodes and a probability for a direct link to exist between two nodes is $p=0.05$.}
\label{fig:MM}
\end{figure}

As explained in the {Appendix}~\ref{sec:mmmod} one can numerically determine the instability region by evaluating the polynomials $S_1(\xi)$ and $S_2(\xi)$ as a function of the model parameters (see panel a) of Fig.~\ref{fig:MM}).  We can observe that the instability region does not intersect the negative real axis and thus only complex eigenvalues, associated to non-symmetric coupling can enter into such region. This is the case reported in panel a) of Fig.~\ref{fig:MM} and confirmed by the time evolution of the density $u_i(t)$ (panel b) in the case of non symmetric long-range interactions and panel c) in the symmetric case). The non-reciprocal interactions are obtained with a directed Erd\H{o}s-R\'enyi network made of $50$ nodes and the probability to create a directed link to be $0.05$. 

\subsection{The Volterra model}
\label{ssec:V}

By remaining in the framework of ecological models, we consider an application dealing with a Volterra model~\cite{mckane2005predator}  {modified by the assumption of the presence of non-reciprocal interactions among animals}. \textcolor{black}{For a sake of pedagogy, we hereby decided to present our results with a} {simplified {and abstract} model} that describes the interactions of {abstract} prey and predators in an ecological setting, \textcolor{black}{let us however observe that our theory could be modified as to include other phenomena, e.g., harvesting~\cite{LayekPati2019,PatiLayekPal2020}. The model we are interesting in results thus be described by}:
\begin{equation}
\label{eq:Voltnospace}
\begin{dcases}
\frac{dx}{dt}=- d x+c_1 xy \\
\frac{dy}{dt}=ry - sy^2 - c_2xy\, .
 \end{dcases}
\end{equation}
Here $x$ denotes the concentration of predators, while $y$ stands for the preys, and all the parameters are assumed to be positive. The Volterra model~\eqref{eq:Voltnospace} admits a nontrivial fixed-point, $x^* = \frac{c_1r-sd}{c_1c_2}$, $y^*=\frac{d}{c_1}$, which is positive and stable, provided that $c_1r-sd>0$.  {Condition hereby assumed to hold true.}

Following the above presented scheme, let us now consider the long-range coupling
\begin{equation*}
 \mathbf{F}(\mathbf{x}^{(i)},\mathbf{x}^{(j)})=\left(-dx_i+ac_1x_jy_i+(1-a)c_1x_iy_j,ry_i-sy_i^2-c_2x_jy_i \right)\, ,
\end{equation*}
where $\mathbf{x}^{(i)}=(x_i,y_i)$ and $\mathbf{x}^{(j)}=(x_j,y_j)$. The parameter $a$ belongs to $[0,1]$ and weights the contribution arising from the nonlinear terms $x_iy_j$ and $x_jy_i$; such term disappears in the local dynamics. Let $\mathbf{f}(\mathbf{x}^{(i)}) = \left(-dx_i+c_1x_iy_i,ry_i-sy_i^2-c_2x_iy_i\right)$, then $ \mathbf{F}(\mathbf{x}^{(i)},\mathbf{x}^{(i)})=\mathbf{f}(\mathbf{x}^{(i)})$. In Appendix~\ref{sec:voltmod} we show that the main equation~\eqref{eq:sysnA} can now be written as
\begin{equation}
\label{eq:VoltnospaceL}
\begin{dcases}
 \frac{dx_i}{dt}&=- d x_i+c_1 x_iy_i+a c_1 y_i\sum_j \mathcal{L}_{ij}x_j +(1-a) c_1 x_i\sum_j \mathcal{L}_{ij}y_j \\
\frac{dy_i}{dt}&=ry_i - sy_i^2- c_2x_iy_i-c_2y_i\sum_j \mathcal{L}_{ij}x_j \, ,
 \end{dcases}
\end{equation}
where we can recognise the $n$ copies of the isolated Volterra system~\eqref{eq:Voltnospace} and the coupling due to the long-range interactions.

\begin{figure}[t]
\includegraphics[scale=.3]{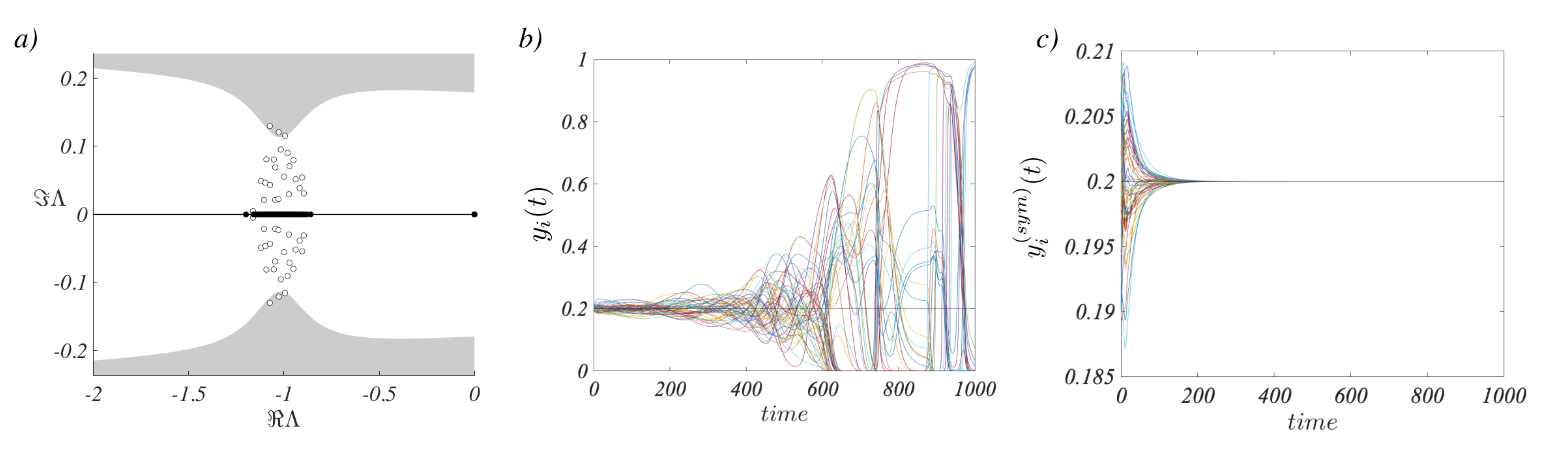}
\caption{\textbf{Instability region and patterns for the Volterra model}. In panel a) we report the region of the complex plane $(\Re \Lambda,\Im \Lambda)$ for which the instability condition is satisfied (grey). For the chosen parameters values ($c_1 = 2$, $c_2= 13$, $r = 1$, $s= 1$, $d = 0.4$ and $a = 0.05$) we can observe that the instability region does not intersect the real axis and thus only non-reciprocal coupling can exhibit complex eigenvalues (white dots) entering into the instability region and thus initiate the pattern as shown in panel b) where we report $y_i(t)$ vs $t$ starting from initial conditions close to the stable equilibrium. Any symmetric coupling determines real eigenvalue (black dots in panel a)) that cannot give rise to the instability as reported in the panel c) where we show $y_i(t)$ vs $t$. The underlying coupling is a directed Erd\H{o}s-R\'enyi network with $n=50$ nodes and a probability for a direct link to exist between two nodes is $p=0.5$.}
\label{fig:patcmplx}
\end{figure}

We can thus compute the explicit form of the polynomials $S_1(\xi)$ and $S_2(\xi)$ as a function of the model parameters (see  {Appendix}~\ref{sec:voltmod}) and characterise the instability region defined by~\eqref{eq:instabcond} as reported in Fig.~\ref{fig:patcmplx} where we show in the complex plane $(\Re \Lambda,\Im \Lambda)$ the regions for which the instability condition is satisfied (grey), for a given set of parameters. By using again a directed Erd\H{o}s-R\'enyi network made of $50$ nodes and the probability to create a directed link to be $0.5$ to describe the non-reciprocal interactions, we can show the existence of eigenvalues $\Lambda^{(\hat{\alpha})}$ (white dots in panel a) of Fig.~\ref{fig:patcmplx}) belonging to this region, associate thus to a patchy solution (panel b) of Fig.~\ref{fig:patcmplx}). The symmetric coupling being associated to a real spectrum is unable to drive the system away from the homogenous equilibrium (panel c) of Fig.~\ref{fig:patcmplx}). Because of the peculiar shape of the instability region, one can have parameters values for which the instability region intersects the real axis (see Fig.~\ref{fig:patcmplxsymm} in {Appendix}~\ref{sec:voltmod}). In this case also a symmetric coupling can trigger the instability.  {From this example one can draw a general conclusion: if the instability is possible by using symmetric long-range interactions, then the same holds true for non-reciprocal ones; on the other hand, patterns resulting from an instability due to non-symmetric interactions can never emerge if the long-range interactions do reciprocate.}

\subsection{The FitzHugh-Nagumo model}
\label{ssec:FHN}

The next system we consider is the FitzHugh-Nagumo model~\cite{fitzhugh,nagumo,Rinzel}, a nonlinear system often used as paradigm for the study of the emergence of Turing patterns~\cite{Ohta,Shoji,CarlettiNakao,siebert,CarlettiNakao} as well as for synchronisation phenomena~\cite{Vragovic,AQIL20121615}. The model has been conceived in the framework of neuroscience as a schematisation of an electric impulse propagating through an axon, for this reason, we believe that it would make a suitable example for the long-range interactions we are describing in this work.

The FitzHugh-Nagumo model can be described by the system of ODE
\begin{equation}
\label{eq:FHN}
\begin{dcases}
\frac{du}{dt}&= \mu u-u^3-v \\
\frac{dv}{dt}&= \gamma(u-\beta v) \, ,
\end{dcases}
\end{equation}
where the parameters $\mu$, $\gamma$ and $\beta$ are assumed to be positive. We will hereby focus on its behaviour close to the fixed point $(u_*,v_*)=(0,0)$. The linear stability analysis ensures stability of the latter under the conditions $\mu<\gamma\beta$ and $\mu\beta<1$. Let us observe that, once such conditions are not met, the system undergoes a supercritical Hopf-Andronov bifurcation~\cite{strogatzbook}: the equilibrium point becomes unstable giving birth to a limit cycle solution. In this study we will limit ourselves to the former case, leaving the oscillating case for a future work. Let us also observe that this model is not a kinetic one, since the $-v$ term appearing in the rate evolution for $u$, expresses a negative cross-effect~\cite{toth}; we can thus claim that pattern formation finds applications beyond morphogenesis and chemical frameworks.

By setting $\mathbf{x}^{(i)}=(u_i,v_i)$ and $\mathbf{f}(\mathbf{x}^{(i)})=(\mu u_i-u_i^3-v_i,\gamma(u_i-\beta v_i))$, we can cast the model in the framework presented in the previous section by using the following coupling
\begin{equation*}
 \mathbf{F}(\mathbf{x}^{(i)},\mathbf{x}^{(j)})=(\mu u_i-u_i^3-v_j,\gamma (u_j-v_i))\, ,
\end{equation*}
where $\mathbf{x}^{(j)}=(u_j,v_j)$, that clearly satisfies $ \mathbf{F}(\mathbf{x}^{(i)},\mathbf{x}^{(i)})=\mathbf{f}(\mathbf{x}^{(i)})$. The main equation~\eqref{eq:sysnA} becomes (see Appendix~\ref{sec:FHNmod})
\begin{equation}
 \label{eq:sysnAFHN}
 \begin{dcases}
\frac{d u_i}{dt} &= \mu u_i-u_i^3-v_i-\sum_j \mathcal{L}_{ij}v_j\\
\frac{d v_i}{dt} &= \gamma(u_i-\beta v_i)+\gamma\sum_j \mathcal{L}_{ij}u_j
\end{dcases} \quad\forall i=1,\dots, n\, ,
\end{equation}
namely a set of $n$ FitzHugh-Nagumo models~\eqref{eq:FHN} coupled via long-range connections in both variables.

\begin{figure}[t]
\includegraphics[scale=.3]{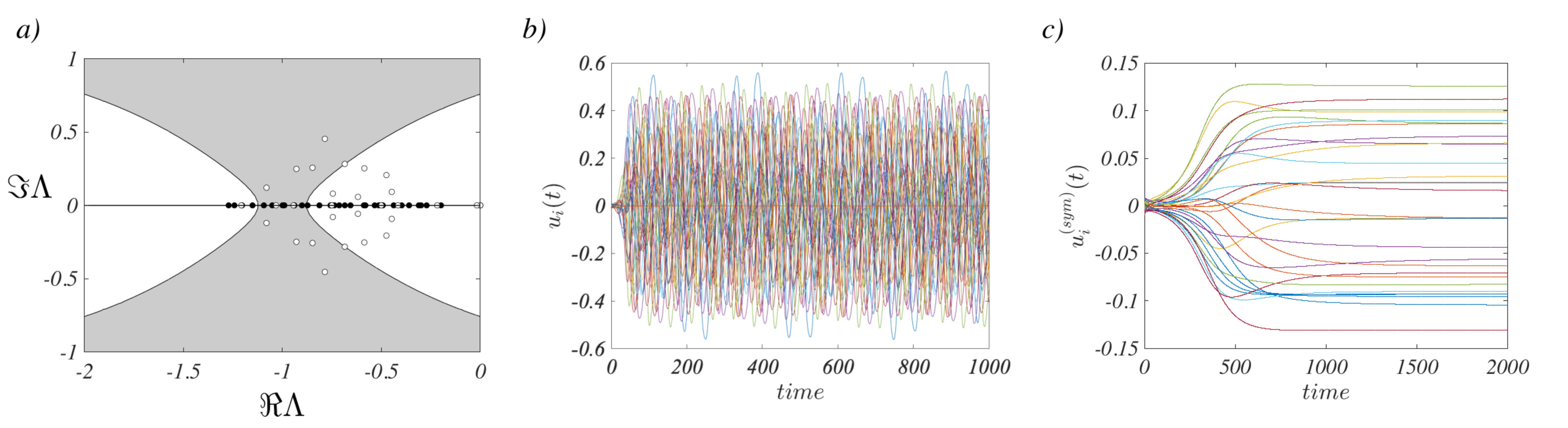}
\caption{\textbf{Instability region and patterns for the FitzHugh-Nagumo model}. In panel a) we report the region of the complex plane $(\Re \Lambda,\Im \Lambda)$ for which the instability condition is satisfied (grey). For the chosen parameters values ($\alpha=1.5$, $\beta=0$, $\gamma=2.5$ and $\mu = 0.01$) we can observe that the instability region intersects the real axis and thus both reciprocal (black dots) and non-reciprocal coupling (white dots) can exhibit  eigenvalues entering into the instability region and thus initiate the pattern as shown in panels b) and c) where we report $u_i(t)$ vs $t$ starting from initial conditions close to the stable equilibrium, $u_*=0$ for the asymmetric coupling and the symmetric one. The underlying coupling is obtained with a directed Erd\H{o}s-R\'enyi network with $n=30$ nodes and a probability for a direct link to exist between two nodes is $p=0.1$.}
\label{fig:FHN}
\end{figure}

We can then compute the polynomials $S_1(\xi)$ and $S_2(\xi)$ (see {Appendix}~\ref{sec:FHNmod}) and determine the instability region (grey) as reported in the panel a) of Fig.~\ref{fig:FHN}. By using a directed Erd\H{o}s-R\'enyi network made of $30$ nodes and the probability to create a directed link to be $0.1$ to encode the non-reciprocal interactions, we can show the existence of eigenvalues (white dots) lying in the instability region and thus allowing for the instability onset (panel b) of Fig.~\ref{fig:FHN}). Let observe that because of the shape of the instability region, exhibiting a non-empty intersection with the negative real axis, there can also be symmetric couplings for which patterns do emerge (panel c) of Fig.~\ref{fig:FHN}).

\subsection{The Stuart-Landau model}
\label{ssec:SL}
As anticipated, the proposed method goes beyond the framework above presented dealing with stationary homogeneous solution, but can be extended to study systems exhibiting an oscillating behaviour, being the latter a regular or a chaotic one. To support this claim, let us study for sake of definitiveness the Stuart-Landau system (SL)~\cite{Stuart1978,Kuramoto}, as paradigmatic model of nonlinear oscillators. The complex amplitude $w$ evolves thus in time according to 
\begin{equation}
\label{eq:SLisol}
\frac{dw}{dt}=\sigma w-\beta |w|^2w\, ,
\end{equation}
where $\sigma=\sigma_\Re+i\sigma_\Im$ and $\beta=\beta_\Re+i\beta_\Im$ are complex parameters, and we denote by $z_\Re$, resp. $z_\Im$, the real, resp. imaginary, part of the complex number $z$. Let us observe that the SL determines the normal form for a generic system close to a supercritical Hopf-bifurcation, hence the results hereby presented are more general than the specific model explored. Let us thus consider a system made of $n$ identical SL oscillators and set the parameters such that each isolated oscillator converges to the same limit cycle  {$w_{LC}(t)=\sqrt{\sigma_\Re/\beta_\Re}e^{i \omega t}$, $\omega=\sigma_\Im-\beta_\Im \sigma_\Re/\beta_\Re$}. We are then interested in determining the conditions responsible for the persistence of this synchronous behaviour once the SL oscillators are allowed to interact through non-reciprocal long-range interactions or, if on the contrary, an instability sets up and drives the whole system to a new (spatially  {possibly}) heterogeneous state. Assume again the long-range interactions to be modelled through a mean field ansatz~\eqref{eq:sysnA}, for instance
\begin{equation}
\label{eq:SL}
\frac{dw_j}{dt}=\frac{\sigma}{k^{(in)}_j}\sum_{\ell} A_{j\ell} w_\ell-\beta w_j|w_j|^2= \sigma w_j-\beta w_j|w_j|^2+\sigma \sum_{\ell} \mathcal{L}_{j\ell} w_\ell \, ,
\end{equation}
where $w_j$ is the complex state variable of the $j$-th SL system and $A_{j\ell}$ encodes the non-reciprocal coupling. Such system admits an homogeneous stable limit cycle solution if $\sigma_\Re>0$ and $\beta_\Re>0$. We can then show the existence of non-reciprocal couplings able to trigger the instability by destabilising the limit cycle solution, eventually driving the system toward a new heterogeneous wavy solution. 

\begin{figure}[t]
\includegraphics[scale=.3]{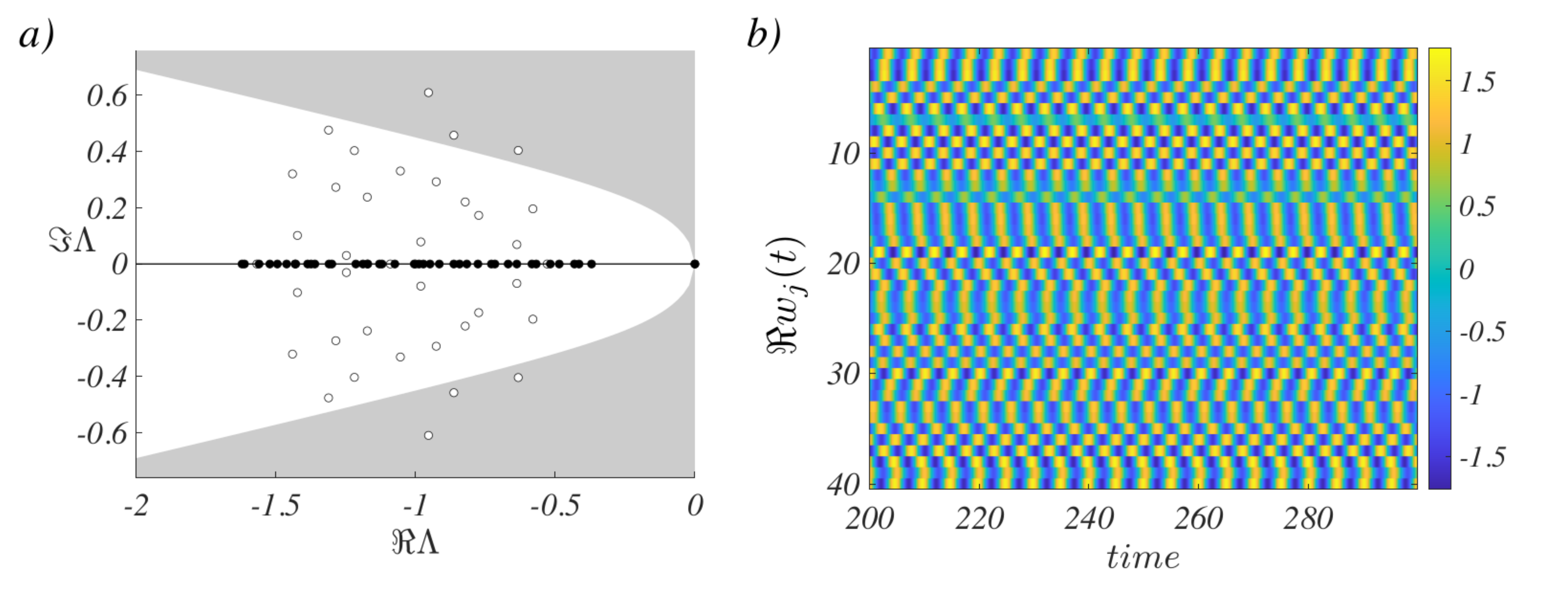}
\caption{\textbf{Instability region and patterns for the Stuart-Landau model}. The region of instability in the complex plane $(\Re \Lambda,\Im \Lambda)$ is reported (grey) in  panel a). We fixed the model parameters to the values $\sigma_\Re = 1$, $\sigma_\Im=4.3$, $\beta_\Re=1$ and $\beta_\Im=1$, resulting into an instability region that does not intersect the real axis, thus only an asymmetric coupling (white dots) can posses a complex spectrum entering the instability region and initiate an heterogenous pattern as shown in panel b) where we report $\Re w_j(t)$ vs time. Any symmetric coupling determines real eigenvalue (black dots in panel a)) and thus it cannot initiate the instability, in consequence all the units do synchronise (data not shown). The underlying coupling is a directed Erd\H{o}s-R\'enyi network with $n=40$ nodes and a probability for a direct link to exist between two nodes is $p=0.08$.}
\label{fig:patcmplxSL}
\end{figure}

Indeed, according to the theory hereby developed, we can always determine model parameters allowing for an instability region in the complex plane (see  {Appendix}~\ref{sec:SLmod}), that does not intersect the real axis; the spectrum of a reciprocal web of long-range interactions could thus never belong to the instability region (black dots in Fig.~\ref{fig:patcmplxSL}) and any perturbation fades away. On the other hand, the complex spectrum associated to non-reciprocal couplings could intersect the instability region (white dots in Fig.~\ref{fig:patcmplxSL}), driving thus the system toward the formation of patterns.  {We can thus state a claim similar to the ones made above, namely non-reciprocal long-range interactions can more easily drive the system toward a desynchronised state.}

\subsection{A sufficient condition for $d\geq 2$}
\label{ssec:dgt2}

 {As already anticipated, the $d=2$ dimensional systems have been presented for sake of pedagogy, because they allow to obtain explicit and analytical conditions for the onset of heterogeneous solutions, where the role of the non-reciprocal links is clearly understood. We are now able to show that similar results hold true in any dimensions, $d\geq 2$, and for generic systems beyond the paradigmatic examples above presented. We will indeed prove that given a system built using reciprocal interactions for which the reference solution is stable and thus any small enough perturbation about it, will fade away, then one can always find non-reciprocal interactions capable to destabilise the reference solution, provided a condition on the model parameters holds true, and can thus drive the system toward a new heterogenous asymptotic state.}

 {The starting point is thus the general system given by~\eqref{eq:sysnAL} and  Eq.~\eqref{eq:finalEqalphaHom} obtained by linearising the dynamics about the stationary reference solution, $\mathbf{s}_0$, and then by projecting onto the Laplace eigenbasis, $\phi^{(\alpha)}$, $\alpha=1,\dots,n$. The characteristic polynomial, whose roots determine the stability feature of the reference solution, is now given by:
\begin{equation}
\label{eq:dispreld}
p_d(\lambda):=(-1)^d\left[ a_0 \lambda^d+a_1\lambda^{d-1}+a_2\lambda^{d-2}+\dots+a_d\right]\, ,
\end{equation}
where the coefficients $a_j$, $j=0,\dots,d$, depend on the model parameters and on the eigenvalues $\Lambda^{(\alpha)}$. A straightforward computation allows to obtain for instance
\begin{equation*}
 a_0=1\,,\quad a_1=-\mathrm{tr}\mathbf{J}^{(\alpha)}\quad \text{and} \quad a_d=\det\mathbf{J}^{(\alpha)}\, ,
\end{equation*}
with $\mathbf{J}^{(\alpha)}=\mathbf{J}(\mathbf{s}_0)+\mathbf{J}_2(\mathbf{s}_0)\Lambda^{(\alpha)}$; it is thus clear that Eq.~\eqref{eq:disprel2} is a particular case of the latter. A more cumbersome computation (see Appendix~\ref{sec:generaldcase}) allows to obtain an explicit formula for $a_2$ that will be needed in the following. In the general $d$-dimensional case, there are not explicit formulas for the roots of the polynomial $p_d$ and even if they exist, their use to understand the role of the model parameters and the impact of the eigenvalues $\Lambda^{(\alpha)}$ will be hopeless. We have thus to resort to a different approach based on the Routh-Hurwitz stability criterion~\cite{Routh1877,Hurwitz1895,Barnett1983}, allowing to prove the (in)stability feature of a real coefficients polynomial. In particular we will use the fact that a necessary condition to have stable polynomial, i.e., a polynomial whose roots have negative real parts, is that all its coefficients exhibit the same sign, in the present case they should be positive being $a_0=1$.
}

 {Before to proceed further, we observe that the coefficients $a_j$ of the polynomial $p_d$ are not (in general) real numbers, because of their dependence on the complex $\Lambda^{(\alpha)}$, one cannot thus directly apply the above criterium. To overcome this issue we introduce the polynomial $q(\lambda)$ of degree $2d$ given by:
\begin{equation*}
q(\lambda) := p_d(\lambda)\overline{p_d(\overline{\lambda})}= \lambda^{2d}+b_1\lambda^{2d-1}+b_2\lambda^{2d-2}+\dots+b_{2d}\, ,
\end{equation*}
where $\overline{z}$ denotes the complex conjugate of the complex number $z$. One can prove (see Appendix~\ref{sec:generaldcase}) that the roots of this polynomial have the same real parts of the roots of $p_d$ (counted twice) and moreover its coefficients become real numbers, whose expression can be related to the coefficients of $p_d$.}

 {The proof of our statement proceeds thus as follows. Assume to have a system whose nodes do interact through a symmetric ensemble of links and fix the model parameters in such a way the reference solution $\mathbf{s}_0$ is stable, with respect to heterogeneous perturbations. Hence for all $\alpha=1,\dots,n$, all the roots $\lambda_i$, $i=1,\dots,d$, of $p_d$ have negative real parts and so do the ones of $q$. Invoking the necessary condition of the Routh-Hurwitz stability criterion we can conclude that the coefficients of $q$ are positive numbers whenever we set $\Im \Lambda^{(\alpha)}=0$, for all $\alpha$, because the Laplace spectrum is real (see Appendix~\ref{sec:spectLapsymm}):
\begin{equation*}
b_j\rvert_{\Im \Lambda^{(\alpha)}=0}>0\quad  \forall j=1,\dots,2d\, .
\end{equation*}
We can then prove (we refer to Appendix~\ref{sec:generaldcase} for more details) that if the model parameters satisfy the condition
\begin{equation}
\label{eq:consign}
\mathrm{tr}\mathbf{J}^2_2<0\, ,
\end{equation}
then we can always find a non-reciprocal coupling such that the imaginary part of the spectrum does satisfy the condition
\begin{equation}
\label{eq:condx}
\exists \hat{\alpha}>1\text{ s.t.}  \left(\Im \Lambda^{(\hat{\alpha})}\right)^2 > \frac{\left(2\Re a_2+\left[\mathrm{tr}\mathbf{J}^{(\hat{\alpha})}\right]^2\right)\rvert_{\Im \Lambda^{(\hat{\alpha})}=0}}{|\mathrm{tr}\mathbf{J}_2^2|} \, ,
\end{equation}
ensuring that the coefficient $b_2$ is negative and thus the polynomial $q$ is not stable. It admits thus at least one root, the one associated to $\hat{\alpha}$, whose real part is positive, and so does $p_d$. In conclusion the reference solution $\mathbf{s}_0$ is unstable, any initial perturbation is amplified and the system moves away from the homogeneous solution. Let us observe that we can always build a web of non-reciprocal interactions satisfying the latter conditions, by using the ideas developed in~\cite{NCFBC} to generate asymmetric Laplacian matrices, and thus directed networks, that possess prescribed spectra.}

\section{Conclusion}
\label{sec:discconcl}
In conclusion, we have proposed and analysed a mechanism for pattern formation, being the latter stationary or time dependent, rooted on the presence of non-reciprocal long-range interactions, to show that self-organisation can manifest without the presence of diffusion. Long-range non-reciprocal interactions, modelled with a mean-field scheme, return a consensus Laplace operator, whose complex spectrum sets the conditions for the instability and the ensuing spatial or temporal patterns. The intrinsic asymmetries within the system, described by non-reciprocal interactions, are the key factor in the disruption of the homogeneous solution and thus the driver for the diversity of patterns of complexity one can observe in Nature. The theory has been applied to several paradigmatic models used in the literature to study the emergence of patterns or synchronisation. The proposed mechanism complements thus the Turing one~\cite{Turing,NM2010} and it is suitable for all phenomena where patterns emergence is not driven by a diffusive process but immobile species interact through non-reciprocal long-range couplings.
\\

\textit{Acknowledgements}\\
We thank Luca Gallo and Jean-François de Kemmeter for discussions. RM is supported by an FRIA-FNRS PhD fellowship, Grant FC 33443, funded by the Walloon region.

\bibliographystyle{apsrev4-1}
\bibliography{bib_RTP}

\newpage

\appendix

\section{About the spectrum of the reactive Laplacian in case of reciprocal interactions.}
\label{sec:spectLapsymm}

Let us assume the network of interactions to be symmetric, $A_{ij}=A_{ji}$, for all $i$ and $j$, and let us recall the definition of the reactive Laplacian $\mathcal{L}_{ij}=A_{ij}/k_i-\delta_{ij}$, where $k_i=\sum_jA_{ij}$, is the node degree. Let us introduce the symmetric Laplace matrix $\mathcal{L}^{sym}_{ij}=A_{ij}/\sqrt{k_ik_j}-\delta_{ij}$ and observe that its eigenvalues are real and negative but the null one. Let $\mathbf{D}$ be the diagonal matrix with the nodes degree on the diagonal, then 
\begin{equation*}
 \mathcal{L} = \mathbf{D}^{-1}\mathbf{A}-\mathbf{I}=\mathbf{D}^{-1/2}\left[\mathbf{D}^{-1/2}\mathbf{A}-\mathbf{I}\right]\mathbf{D}^{1/2}=\mathbf{D}^{-1/2}\mathcal{L}^{sym}\mathbf{D}^{1/2}\, .
\end{equation*}

Namely $\mathcal{L}$ and $\mathcal{L}^{sym}$ are similar matrices and thus exhibit the same set of eigenvalues.

\section{Linear stability analysis}
\label{sec:dynb}

Let us consider the homogeneous reference solution $\mathbf{s}(t)$ and a spatially dependent perturbation about the latter, $\mathbf{x}^{(i)}(t)=\mathbf{s}(t)+\mathbf{u}^{(i)}(t)$, then by inserting this information into Eq.~\eqref{eq:sysnA} (main text) and by retaining only the linear terms in $\mathbf{u}^{(i)}$ we obtain
\begin{eqnarray*}
 \frac{d\mathbf{u}^{(i)}}{dt}= \frac{d\mathbf{x}^{(i)}}{dt}-\frac{d\mathbf{s}}{dt}&=&\frac{1}{k^{(in)}_i}{\sum_j A_{ij} \mathbf{F}(\mathbf{s}+\mathbf{u}^{(i)},\mathbf{s}+\mathbf{u}^{(j)})} - \mathbf{f}(\mathbf{s})\\&=&\frac{1}{k^{(in)}_i}{\sum_j A_{ij} \left(\sum_{\ell}\partial_{x_{\ell}^{(i)}}\mathbf{F}(\mathbf{s},\mathbf{s}){u}_{\ell}^{(i)}+\sum_{\ell}\partial_{{x}_{\ell}^{(j)}}\mathbf{F}(\mathbf{s},\mathbf{s}){u}_{\ell}^{(j)}\right)}\\
 &=&\sum_{\ell}\partial_{x_{\ell}^{(i)}}\mathbf{F}(\mathbf{s},\mathbf{s}){u}_{\ell}^{(i)}+\frac{1}{k^{(in)}_i}{\sum_j A_{ij} \sum_{\ell}\partial_{{x}_{\ell}^{(j)}}\mathbf{F}(\mathbf{s},\mathbf{s}){u}_{\ell}^{(j)}}\\
 &=&\mathbf{J}_1\mathbf{u}^{(i)}+\frac{1}{k^{(in)}_i}{\sum_j A_{ij} \mathbf{J}_2\mathbf{u}^{(j)}} \quad\forall i=1,\dots, n\, ,
\end{eqnarray*}
where we recall that  {$\mathbf{J}_1=\partial_{\mathbf{x}_1} \mathbf{F}(\mathbf{x}_1,\mathbf{x}_2)_{(\mathbf{s},\mathbf{s})}$} and  {$\mathbf{J}_2=\partial_{\mathbf{x}_2} \mathbf{F}(\mathbf{x}_1,\mathbf{x}_2)_{(\mathbf{s},\mathbf{s})}$}. By observing that $\mathbf{f}(\mathbf{x})=\mathbf{F}(\mathbf{x},\mathbf{x})$ one can prove that $\partial_{\mathbf{x}}\mathbf{f}:= \mathbf{J}=\mathbf{J}_1+\mathbf{J}_2$. Hence by slightly rewriting the previous equation, we obtain
\begin{eqnarray}
\label{eq:finalEq}
 \frac{d\mathbf{u}^{(i)}}{dt}(t)&=&\mathbf{J}_1\mathbf{u}^{(i)}+\frac{1}{k^{(in)}_i}{\sum_j A_{ij} \mathbf{J}_2\mathbf{u}^{(j)}}=\mathbf{J}_1\mathbf{u}^{(i)}+\mathbf{J}_2\mathbf{u}^{(i)}+\sum_j \left(\frac{A_{ij}}{k^{(in)}_i}-\delta_{ij}\right) \mathbf{J}_2\mathbf{u}^{(j)}\notag \\&=&\left(\mathbf{J}_1+\mathbf{J}_2\right)\mathbf{u}^{(i)}+\sum_j \mathcal{L}_{ij} \mathbf{J}_2\mathbf{u}^{(j)}\notag \\&=&\mathbf{J}\mathbf{u}^{(i)}+\sum_j \mathcal{L}_{ij} \mathbf{J}_2\mathbf{u}^{(j)} \, ,
\end{eqnarray}
where we used the matrix $\mathcal{L}$, given by Eq.~\eqref{eq:consLapHG} (main text).

By introducing the $n\times d$ vector $\mathbf{u}=((\mathbf{u}^{(1)})^\top,\dots,(\mathbf{u}^{(n)})^\top)^\top$, we can eventually rewrite the latter equation in a compact form as:
\begin{equation}
 \frac{d\mathbf{u}}{dt}(t)=\left[ \mathbf{I}_n\otimes \left(\mathbf{J}_1+\mathbf{J}_2\right)+ \mathcal{L} \otimes \mathbf{J}_2\right]\mathbf{u} = \left[ \mathbf{I}_n\otimes\mathbf{J}+ \mathcal{L} \otimes \mathbf{J}_2\right]\mathbf{u}\, ,
\label{eq:lincompact}
\end{equation}
where $\otimes$ denotes the Kronecker product of matrices. In this way we emphasise the  role of the Jacobian of the isolated system $\mathbf{I}_n\otimes\mathbf{J}$ and the one for the coupling part, $\mathcal{L} \otimes \mathbf{J}_2$, that vanishes once we assume $\mathbf{A}=\mathbf{I}_n$.

Eq.~\eqref{eq:finalEq} is a linear system involving matrices with size $nd\times nd$. One can reduce the complexity of the latter by assuming the existence of an eigenbasis of $\mathcal{L}$, $\phi^{(\alpha)}$, $\alpha=1,\dots, n$, with associated eigenvalue $-2<\Re\Lambda^{(\alpha)}\leq 0$. Then by rewriting $\mathbf{u}^{(i)}=\sum_\beta \mathbf{u}^{(\beta)}\phi^{(\beta)}_i$ and inserting the latter into~\eqref{eq:finalEq} we obtain
\begin{eqnarray*}
\sum_\beta \frac{d\mathbf{u}^{(\beta)}}{dt}\phi_i^{(\beta)}&=&\sum_\beta\mathbf{J}(\mathbf{s}(t))\mathbf{u}^{(\beta)}\phi_i^{(\beta)}+\sum_j\sum_\beta \mathcal{L}_{ij}\mathbf{J}_2 (\mathbf{s}(t))\mathbf{u}^{(\beta)}\phi_j^{(\beta)}\\
&=&\sum_\beta\mathbf{J}(\mathbf{s}(t))\mathbf{u}^{(\beta)}\phi_i^{(\beta)}+\sum_\beta \Lambda^{(\beta)}\mathbf{J}_2 (\mathbf{s}(t))\mathbf{u}^{(\beta)}\phi_i^{(\beta)}\, .
\end{eqnarray*}
 By multiplying by the left eigenvectors $\psi_i^{(\alpha)}$ and summing over the index $i$, recall that $\sum_i\psi_i^{(\alpha)}\phi_i^{(\beta)}=\delta_{\alpha\beta}$, we can obtain Eq.~\eqref{eq:finalEqalpha} (main text), namely
\begin{equation*}
\frac{d\mathbf{u}^{(\alpha)}}{dt}=\mathbf{J}(\mathbf{s}(t))\mathbf{u}^{(\alpha)}+ \Lambda^{(\alpha)}\mathbf{J}_2 (\mathbf{s}(t))\mathbf{u}^{(\alpha)}\quad \forall \alpha=1,\dots,n\, .
\end{equation*}

Assuming to deal with $d=2$ dimensional systems and with a stationary reference solution, i.e., $\mathbf{s}(t)=\mathbf{s}_0$ for all $t$; one can then realise that the eigenvalues of  $\mathbf{J}^{(\alpha)}$, $\lambda_i$, are the solutions of the second order equation
\begin{equation*}
 \lambda_i^2-\mathrm{tr}\mathbf{J}^{(\alpha)}\lambda_i+\det \mathbf{J}^{(\alpha)}=0\quad \forall i=1,\dots, n\, .
\end{equation*}
Note that if on the other hand, we were interested in studying $d>2$ dynamical systems, then we should consider roots of $d$-degree polynomial equations for which there is no general closed form solution and thus one has to recur to numerics to determine the instability region  {or to another ideas as we have shown in the main text and detailed in Appendix~\ref{sec:generaldcase}}.

Back to the $d=2$ case, one can express the real part of the above roots as:
\begin{equation*}
\Re\lambda_i = \frac{1}{2}\left(\Re \mathrm{tr}\mathbf{J}^{(\alpha)}+\gamma\right)\, ,
\end{equation*}
where
\begin{equation*}
 \gamma=\sqrt{\frac{A+\sqrt{A^2+B^2}}{2}}\, , A=\left(\Re\mathrm{tr}\mathbf{J}^{(\alpha)}\right)^2-\left(\Im\mathrm{tr}\mathbf{J}^{(\alpha)}\right)^2-4\Re\mathrm{det}\mathbf{J}^{(\alpha)}\text{ and }B=2\Re\mathrm{tr}\mathbf{J}^{(\alpha)}\Im\mathrm{tr}\mathbf{J}^{(\alpha)}-4\Im\mathrm{det}\mathbf{J}^{(\alpha)}\, .
\end{equation*}
A straightforward but lengthy computation allows to rewrite the condition for instability Eq.~\eqref{eq:instabcond} (main text), in terms of two polynomials, $S_2$ of second degree and $S_1$ of fourth degree. More precisely, $S_2(\xi)=c_{2,2}\xi^2+c_{2,1}\xi+c_{2,0}$ with coefficients
\begin{eqnarray}
\label{eq:cS2}
 c_{2,2}&=& - \det\mathbf{J}_2 (4 \det\mathbf{J}_2 - \left(\mathrm{tr}\mathbf{J}_2\right)^2)\notag\\
 c_{2,1}&=&  -\Delta_1 (4 \det\mathbf{J}_2 - \left(\mathrm{tr}\mathbf{J}_2\right)^2)\\
 c_{2,0}&=& -  \Delta_1^2 + \Delta_1\mathrm{tr}\mathbf{J}\mathrm{tr}\mathbf{J}_2 - \det\mathbf{J}_2 \left(\mathrm{tr}\mathbf{J}\right)^2\notag\, ,
\end{eqnarray}
and $S_1(\xi)=c_{1,4}\xi^4+c_{1,3}\xi^3+c_{1,2}\xi^2+c_{1,1}\xi+c_{1,0}$ with coefficients
\begin{eqnarray}
\label{eq:cS1}
 c_{1,4}&=& \det\mathbf{J}_2 \left(\mathrm{tr}\mathbf{J}_2\right)^2\notag\\
 c_{1,3}&=&  \mathrm{tr}\mathbf{J}_2 (\Delta_1\mathrm{tr}\mathbf{J}_2+ 2 \det\mathbf{J}_2\mathrm{tr}\mathbf{J})\notag\\
 c_{1,2}&=&  \det\mathbf{J} \left(\mathrm{tr}\mathbf{J}_2\right)^2 + 2 \Delta_1\mathrm{tr}\mathbf{J}\mathrm{tr}\mathbf{J}_2 + \det\mathbf{J}_2 \left(\mathrm{tr}\mathbf{J}\right)^2\\
 c_{1,1}&=&  \mathrm{tr}\mathbf{J} (2 \det\mathbf{J} \mathrm{tr}\mathbf{J}_2 + \Delta_1 \mathrm{tr}\mathbf{J})\notag\\
 c_{1,0}&=&  \det\mathbf{J} \left(\mathrm{tr}\mathbf{J}\right)^2\notag\, ,
\end{eqnarray}
where we introduced $\Delta_1=J_{2,11}J_{22} - J_{2,12}J_{21} - J_{2,21}J_{12} + J_{2,22}J_{11}$.\vspace{1cm}

Let us now consider the general case in which the existence of an eigenbasis of $\mathcal{L}$ is not guaranteed, namely the Laplace matrix is defective. We can then invoke the Jordan canonical form theorem to determine an invertible $n\times n$ matrix $\mathbf{P}$ such that
\begin{equation*}
 \mathbf{P}^{-1}\mathcal{L}\mathbf{P}=\mathbf{B}=\mathrm{diag}(\mathbf{B}_1,\dots,\mathbf{B}_\ell)\, ,
\end{equation*}
where the $\mathbf{B}_j$ is the $m_j\times m_j$ Jordan block, $m_1+\dots+m_\ell=n$,
\begin{equation*}
\mathbf{B}_j=\left(
\begin{matrix}
 \Lambda^{(j)} & & & \\
 1&  \Lambda^{(j)}&  & \\
  &   \ddots& \ddots &\\
   & & 1& \Lambda^{(j)}
\end{matrix}
\right)\, ,
\end{equation*}
being $\Lambda^{(j)}$ the $j$-th eigenvalue of $\mathcal{L}$. Because $ \Lambda^{(1)}=0$ we also have $\mathbf{B}_1=0$.

Let us consider again Eq.~\eqref{eq:lincompact}. By defining $\mathbf{Q}=\mathbf{P}\otimes \mathbf{I}_d$ and $\mathbf{v}=\mathbf{Q}^{-1}\mathbf{u}$ we get
\begin{equation}
 \frac{d\mathbf{v}}{dt}(t) =\mathbf{Q}^{-1}\frac{d\mathbf{u}}{dt}(t)=(\mathbf{P}^{-1}\otimes \mathbf{I}_d) \left[ \mathbf{I}_n\otimes\mathbf{J}+ \mathcal{L} \otimes \mathbf{J}_2\right](\mathbf{P}\otimes \mathbf{I}_d)\mathbf{Q}^{-1}\mathbf{u}=\left[ \mathbf{I}_n\otimes\mathbf{J}+\mathbf{B} \otimes \mathbf{J}_2\right]\mathbf{v}\, .
\label{eq:lincompactJordan}
\end{equation}
The vector $\mathbf{v}$ inherits the Jordan decomposition, hence we can write $\mathbf{v}=((\mathbf{v}^{(1)})^\top,\dots,(\mathbf{v}^{(\ell)})^\top)^\top$, where $\mathbf{v}^{(j)}$ is a $(d\times m_j)$-dimensional vector. Eq.~\eqref{eq:lincompactJordan} can thus be rewritten as
\begin{eqnarray}
 \frac{d\mathbf{v}^{(1)}}{dt}(t) &=&\left(
\begin{matrix}
 \mathbf{J} & &\\
 & \ddots &\\
 & &  \mathbf{J}
\end{matrix}\right)\mathbf{v}^{(1)} \label{eq:lincompactJordan1}\, ,\\
 \frac{d\mathbf{v}^{(j)}}{dt}(t) &=&\left(
\begin{matrix}
 \mathbf{J} & &\\
 & \ddots &\\
 & &  \mathbf{J}
\end{matrix}\right)\mathbf{v}^{(j)}+\left(
\begin{matrix}
 \Lambda^{(j)} \mathbf{J}_2& & & \\
 \mathbf{J}_2&  \Lambda^{(j)}\mathbf{J}_2&  & \\
  &   \ddots& \ddots &\\
   & & \mathbf{J}_2& \Lambda^{(j)}\mathbf{J}_2
\end{matrix}
\right)\mathbf{v}^{(j)} \quad \forall j=2,\dots, \ell \, .
\label{eq:lincompactJordanj}
\end{eqnarray}

The first Eq.~\eqref{eq:lincompactJordan1} returns again the stability condition for the isolated systems, which is a working assumption for the developed framework. Let us now consider the generic Eq.~\eqref{eq:lincompactJordanj} and by writing $\mathbf{v}^{(j)}=((\mathbf{v}^{(j)}_1)^\top,\dots,(\mathbf{v}^{(j)}_{m_j})^\top)^\top$, where $\mathbf{v}^{(j)}_i\in\mathbf{R}^d$ for all $i=1,\dots,m_j$, we get
\begin{eqnarray}
 \frac{d\mathbf{v}^{(j)}_1}{dt}(t) &=&\mathbf{J}\mathbf{v}^{(j)}_1+\Lambda^{(j)} \mathbf{J}_2\mathbf{v}^{(j)}_1 \label{eq:lincompactJordanj1}\, , \\
 \frac{d\mathbf{v}^{(j)}_2}{dt}(t) &=&\mathbf{J}\mathbf{v}^{(j)}_2+\Lambda^{(j)} \mathbf{J}_2\mathbf{v}^{(j)}_2 +\mathbf{J}_2\mathbf{v}^{(j)}_1 \label{eq:lincompactJordanj2}\, , \\
& \vdots & \notag\\
\frac{d\mathbf{v}^{(j)}_{m_j}}{dt}(t) &=&\mathbf{J}\mathbf{v}^{(j)}_{m_j}+\Lambda^{(j)} \mathbf{J}_2\mathbf{v}^{(j)}_{m_j} +\mathbf{J}_2\mathbf{v}^{(j)}_{m_j-1} \label{eq:lincompactJordanjmj}\, .
\end{eqnarray}

To simplify the following analysis, we will assume to deal with a stationary reference solution $\mathbf{s}(t)=\mathbf{s}_0$, hence all the involved matrices are constant ones. Hence the first Eq.~\eqref{eq:lincompactJordanj1} is the analogous of Eq.~\eqref{eq:finalEqalpha} (main text) and one can determine a condition on $\Lambda^{(j)}$ to make the projection $\mathbf{v}_1^{(j)}$ unstable; in the case $d=2$ this accounts to perform the analysis above presented returning the condition~\eqref{eq:instabcond}, where $\hat{\alpha}$ is replaced by $j$. 

Let us now consider the second Eq.~\eqref{eq:lincompactJordanj2} and observe that it is composed by two terms, the first one involves the same matrix of the first equation, $\mathbf{J}+\Lambda^{(j)} \mathbf{J}_2$, while the second one depends on the projection $\mathbf{v}_1^{(j)}(t)$. If, for the choice of $\Lambda^{(j)}$ the matrix is unstable and thus $\mathbf{v}_1^{(j)}(t)$ has an exponential growth, then the same is true for $\mathbf{v}_2^{(j)}(t)$. By considering the remaining equations and by exploiting the peculiar shape of the system, this allows to prove that if the first Eq.~\eqref{eq:lincompactJordanj1} returns an unstable solution, then all the solutions $\mathbf{v}_i^{(j)}(t)$ are unstable as well.

In conclusion if the Laplace matrix $\mathcal{L}$ is defective, one can check the instability condition on the available eigenvalues and conclude about the emergence of patterns solely based on this information. Let us observe that this is a sufficient condition, indeed it can happen that the matrix $\mathbf{J}+\Lambda^{(j)} \mathbf{J}_2$ is stable, but the presence of Jordan blocks introduces a transient (polynomial) growth in the linear regime that results strong enough to limit the validity of the linear approximation. Thus the nonlinear system could exhibit orbits departing from the homogeneous reference solution, only infinitesimal perturbations will be attracted to the latter. Stated differently, the stability basin of the reference solution shrinks considerably in presence of defective Laplace matrix, the solution is thus stable but finite perturbations can be amplified.

\section{Analysis of the Brusselator model}
\label{sec:bxlmod}

Let us consider the two species Brusselator model, $\mathbf{f}(\mathbf{x}^{(i)})=(1-(b+1)u_i+c u_i^2v_i,bu_i-c u_i^2v_i)$, where $\mathbf{x}^{(i)}=(u_i,v_i)$, and the non-local coupling given by
\begin{equation*}
 \mathbf{F}(\mathbf{x}^{(i)},\mathbf{x}^{(j)})=(1-(b+1)u_i+c u_i^2v_i,bu_j-c u_i^2v_i)\, ,
\end{equation*}
where $\mathbf{x}^{(j)}=(u_j,v_j)$. Eq.~\eqref{eq:sysnA} can thus be rewritten as
\begin{equation*}
 \begin{dcases}
\frac{d u_i}{dt} &= \frac{1}{k_i^{(in)}}\sum_j A_{ij} \left( 1-(b+1)u_i+c u_i^2v_i\right)\\
\frac{d v_i}{dt} &= \frac{1}{k_i^{(in)}}\sum_j A_{ij}\left( bu_j-c u_i^2v_i\right)
\end{dcases} \quad\forall i=1,\dots, n\, ,
\end{equation*}
by recalling the definition of $k_i^{(in)}=\sum_j A_{ij}$ and by observing that the first equation does not depend on the index $j$, it can be straightforwardly simplified to return
\begin{equation*}
\frac{d u_i}{dt}= 1-(b+1)u_i+c u_i^2v_i\quad\forall i=1,\dots, n\, .
\end{equation*}
Let us now consider the second equation and isolate the terms not depending on the index $j$
\begin{equation*}
\frac{d v_i}{dt} = -c u_i^2v_i+\frac{b}{k_i^{(in)}}\sum_j A_{ij}u_j=-c u_i^2v_i+b\sum_j \left(\frac{A_{ij}}{k_i^{(in)}}-\delta_{ij}\right)u_j + bu_i=bu_i-c u_i^2v_i+b\sum_j \mathcal{L}_{ij}u_j\quad\forall i=1,\dots, n\, .
\end{equation*}
In conclusion we obtain
\begin{equation*}
 \begin{dcases}
\frac{d u_i}{dt} &= 1-(b+1)u_i+c u_i^2v_i\\
\frac{d v_i}{dt} &= bu_i-c u_i^2v_i+b\sum_j \mathcal{L}_{ij}u_j
\end{dcases} \quad\forall i=1,\dots, n\, ,
\end{equation*}

From the definitions of $\mathbf{f}$ and $\mathbf{F}$ we can compute the associated Jacobian matrices:
\begin{equation*}
 \mathbf{J}_{\mathrm{Bxl}}=\left(
\begin{matrix}
b-1 & c\\ -b &-c
\end{matrix}
\right)\text{ and } \mathbf{J}_2=\left(
\begin{matrix}
0 & 0\\ b &0
\end{matrix}
\right)\, ,
\end{equation*}
and thus, by using the formulas~\eqref{eq:cS2} and~\eqref{eq:cS1} for the coefficients $c_{1,i}$, $i=0,1,2,3,4$, and $c_{2,i}$, $i=0,1,2$, we obtain the polynomials $S_1$ and $S_2$:
\begin{equation*}
S_1(\xi) =c(b-c-1)^2(1-b\xi)\text{ and }  S_2(\xi) = -b^2c^2 \, .
\end{equation*}
The condition the complex eigenvalues $\Lambda^{(\alpha)}$ of $\mathcal{L}$ have to satisfy to induce an instability is eventually given by
\begin{equation*}
(\Im \Lambda^{(\alpha)})^2 \geq \frac{(b-c-1)^2}{cb^2} \left(1-b\Re \Lambda^{(\alpha)}\right)\, .
\end{equation*}

\section{Analysis of the Mimura-Murray model}
\label{sec:mmmod}

Let us consider the Mimura-Murray model with reaction terms, $\mathbf{f}(\mathbf{x}^{(i)})=\left(u_i\left( \frac{a+bu_i-u_i^2}{c} - v_i\right),v_i\left( u_i - (1+dv_i) \right)\right)$, where $\mathbf{x}^{(i)}=(u_i,v_i)$. The chosen non-local coupling is
\begin{equation*}
 \mathbf{F}(\mathbf{x}^{(i)},\mathbf{x}^{(j)})=\left(u_i\left( \frac{a+bu_j-u_i^2}{c} - v_j\right),v_i\left( u_j - (1+dv_i) \right) \right)\, .
\end{equation*}
From Eq.~\eqref{eq:sysnA} we obtain
\begin{equation*}
 \begin{dcases}
\frac{d u_i}{dt} &= \frac{1}{k_i^{(in)}}\sum_j A_{ij} u_i\left( \frac{a+bu_j-u_i^2}{c} - v_j\right)\\
\frac{d v_i}{dt} &= \frac{1}{k_i^{(in)}}\sum_j A_{ij}v_i\left( u_j - (1+dv_i) \right)
\end{dcases} \quad\forall i=1,\dots, n\, ,
\end{equation*}
By using again the definition of $k_i^{(in)}$ and by isolating the terms independent from $j$
\begin{equation*}
 \begin{dcases}
\frac{d u_i}{dt} &= u_i\frac{a-u_i^2}{c}+  \frac{u_i}{k_i^{(in)}}\sum_j A_{ij}\left( \frac{u_j}{c} - v_j\right)\\
\frac{d v_i}{dt} &= -v_i(1+dv_i)+\frac{v_i}{k_i^{(in)}}\sum_j A_{ij}u_j 
\end{dcases} \quad\forall i=1,\dots, n\, .
\end{equation*}
By adding and removing suitable terms we eventually obtain
\begin{equation*}
 \begin{dcases}
\frac{d u_i}{dt} &= u_i\left( \frac{a+bu_i-u_i^2}{c} - v_i\right)+\frac{b}{c}u_i\sum_j \mathcal{L}_{ij}u_j-u_i\sum_j \mathcal{L}_{ij}v_j\\
\frac{d v_i}{dt} &= v_i\left( u_i - (1+dv_i) \right)+v_i\sum_j \mathcal{L}_{ij}u_j
\end{dcases} \quad\forall i=1,\dots, n\, .
\end{equation*}

The Jacobian matrices follow directly from the definitions of the functions $\mathbf{f}$ and $\mathbf{F}$:
\begin{equation*}
 \mathbf{J}_{\mathrm{MM}}=\left(
\begin{matrix}
u_*\frac{b-2u_*}{c} & -u_*\\ v_* &-dv_*
\end{matrix}
\right)\text{ and } \mathbf{J}_2=\left(
\begin{matrix}
\frac{b}{c}u_*  & -u_*\\ v_* &0
\end{matrix}
\right)\, ,
\end{equation*}
and from the latter, one can compute the coefficients $c_{1,i}$, $i=0,1,2,3,4$, and $c_{2,i}$, $i=0,1,2$, and eventually the polynomials $S_1$ and $S_2$. In the present case the resulting expressions are quite involved and not very explicative, however they can be used to numerically determine the instability region in the complex plane.

\section{Analysis of the Volterra model}
\label{sec:voltmod}

Let us study a set of $n$ coupled Volterra model~\cite{mckane2005predator} that describes the interactions of preys and predators in an ecological setting where they can interact via long-range connections modelled by
\begin{equation*}
\begin{dcases}
 \dfrac{dx_i}{dt}=- d x_i+a c_1 y_i \frac{1}{k^{(in)}_i}\sum_j A_{ij}x_j +(1-a) c_1 x_i  \frac{1}{k^{(in)}_i}\sum_j A_{ij} y_j \\
 \dfrac{dy_i}{dt}=ry_i - sy_i^2-c_2y_i \frac{1}{k^{(in)}_i}\sum_j A_{ij}x_j \, .
 \end{dcases}
\end{equation*}
By using the newly introduced Laplace matrix~\eqref{eq:consLapHG} we can rewrite the previous equations as:
\begin{equation}
\label{eq:VoltnospaceA2}
\begin{dcases}
 \dfrac{dx_i}{dt}=- d x_i+c_1 y_i x_i +a c_1 y_i \sum_j \mathcal{L}_{ij}x_j +(1-a) c_1 x_i  \sum_j \mathcal{L}_{ij} y_j \\
 \dfrac{dy_i}{dt}=ry_i - sy^2_i - c_2y_i x_i- c_2y_i \sum_j \mathcal{L}_{ij}x_j \, ,
 \end{dcases}
\end{equation}
where one can easily recognise the in-node Volterra model~\eqref{eq:Voltnospace} and the corrections stemming from non-local contributions. The Jacobian matrices are obtained as:
\begin{equation*}
 \mathbf{J}_{\mathrm{V}}=\left(
\begin{matrix}
0 & c_1x^*\\ -c_2y^* &-sy^*
\end{matrix}
\right)\text{ and } \mathbf{J}_2=\left(
\begin{matrix}
ac_1y^*  & (1-a)c_1x^*\\ -c_2y^* &0
\end{matrix}
\right)\, ,
\end{equation*}
where $(x^*,y^*)$ is the homogeneous equilibrium of the Volterra system. 

By inserting the given expressions for $\mathbf{J}_{\mathrm{V}}$ and $\mathbf{J}_2$, in the general formulas~\eqref{eq:cS2} and~\eqref{eq:cS1} we obtain for the coefficients of $S_2(\xi)$
\begin{eqnarray*}
 c_{2,2}&=& -(a-1)(c_1r-ds)\left(4\frac{(a-1)(c_1r-ds)}{c_1}+a^2d\right)\\
 c_{2,1}&=&  (c_1r(a-2)+2ds)\left(4\frac{(a-1)(c_1r-ds)}{c_1}+a^2d\right)\\
 c_{2,0}&=& - \frac{(c_1r(a-2)+2ds)^2}{c_1} +  \left(\frac{c_1r(a-2)+2ds}{c_1}\right){ads} +(a-1)(c_1r-ds)a^2d\, ,
\end{eqnarray*}
and $S_1(\xi)$
\begin{eqnarray*}
 c_{1,4}&=& -(a-1)(c_1r-ds)a^2d\\
 c_{1,3}&=&  ad \left(-(c_1r(a-2)+2ds) a+2\frac{s(a-1)(c_1r-ds)}{c_1}\right)\\
 c_{1,2}&=&  (c_1r-ds)a^2d+2\frac{c_1r(a-2)+2ds}{c_1}ads-\frac{(a-1)(c_1r-ds)}{c_1^2}ds^2\\
 c_{1,1}&=&  -ds\left(2 \frac{c_1r-ds}{c_1}a+\frac{c_1r(a-2)+2ds}{c_1}\frac{s}{c_1}\right)\\
 c_{1,0}&=&  \frac{c_1r-ds}{c_1^2}ds^2\, .
\end{eqnarray*}

\begin{figure}[t]
\includegraphics[scale=.24]{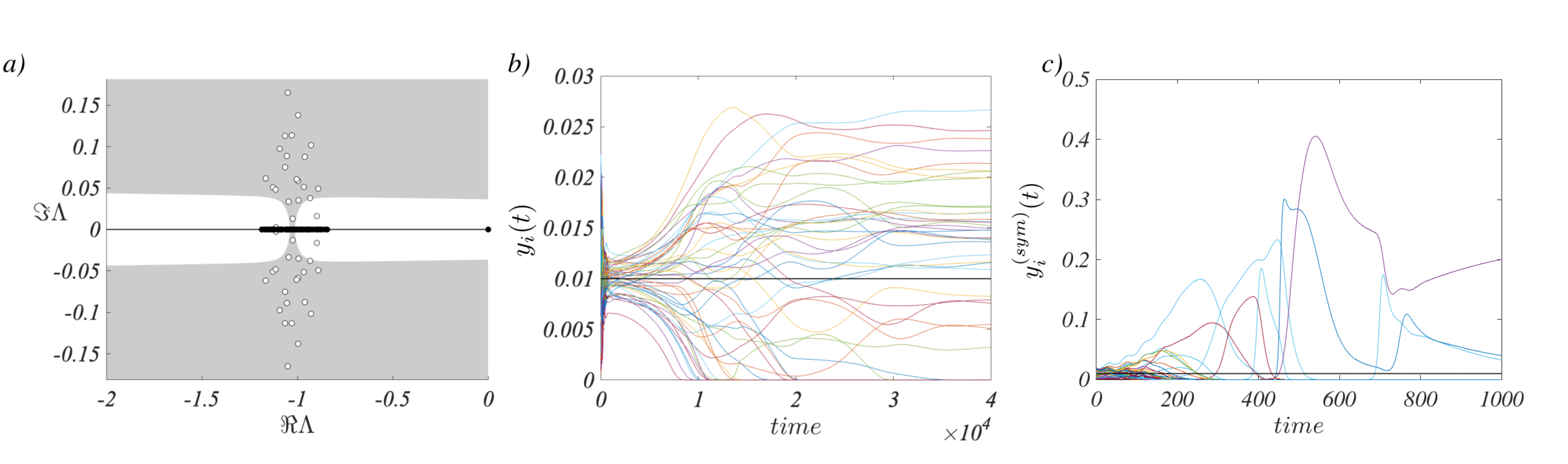}
\caption{\textbf{Instability region and patterns for the Volterra model}. We report the region of the complex plane $(\Re \Lambda,\Im \Lambda)$ for which the instability condition is satisfied (grey), the instability is at play if at least one eigenvalue of $\mathcal{L}$  belongs to the region. The model parameters have been fixed to the values $c_1 = 2$, $c_2= 13$, $r = 1$, $s= 1$, $d = 0.02$ and $a = 0.05$, and we can observe that the instability region intersects the real axis and thus both non-reciprocal (white dots) and reciprocal (black dots) interactions can exhibit eigenvalues entering into the instability region. This in turns implies the existence of an heterogeneous solution for both the reciprocal (see panel b) where we report the density of preys vs time) and non-reciprocal (see panel c) where we report the density of preys vs time) long-range interactions assumption. In both panels the horizontal black line denotes the homogeneous equilibrium $y^*$.}
\label{fig:patcmplxsymm}
\end{figure}

Given such polynomials one can determine the (in)-stability region as shown in Fig.~\ref{fig:patcmplx} (main text) or Fig.~\ref{fig:patcmplxsymm} and thus conclude about the onset of the instability according to the position of the complex eigenvalues of the Laplace matrix $\mathcal{L}$. In Fig. ~\ref{fig:patcmplxsymm} we report the region of instability (grey) for a set of parameters values allowing for the emergence of patterns for both the reciprocal and non-reciprocal long-range interactions. We can observe that, contrary to the case shown in the main text, now the instability region has a non empty intersection with the real axis where it lies the spectrum of the Laplace operator for reciprocal interactions (black dots). We can thus determine a web of symmetrical long-range interactions for which the Volterra model~\eqref{eq:Voltnospace} (main text) exhibits an instability and eventually evolves toward a patchy solution (panel b) of Fig.~\ref{fig:patcmplxsymm}). A similar result holds true using non-reciprocal long-range interactions, indeed the complex spectrum of the associated Laplace matrix (white dots in panel a) of Fig.~\ref{fig:patcmplxsymm}) also lies in the instability region and thus the system converges to a spatially heterogeneous solution (panel c) of Fig.~\ref{fig:patcmplxsymm}). The underlying long-range coupling is a directed Erd\H{o}s-R\'enyi network with $n=50$ nodes and a probability for a direct link to exist between two nodes is $p=0.5$.

\begin{figure}[t]
\includegraphics[scale=.30]{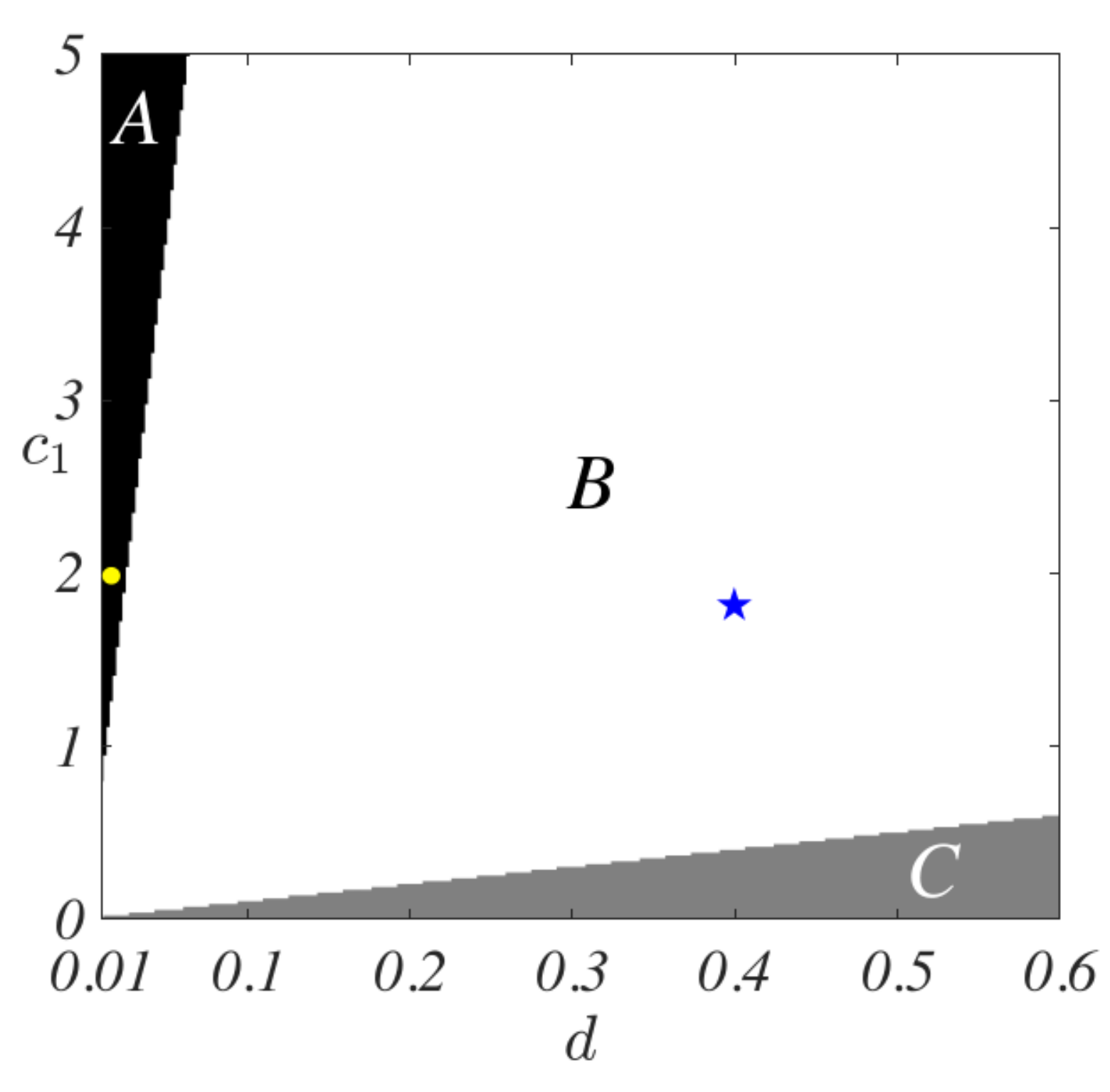}
\caption{\textbf{Bifurcation diagram for the Volterra model, reciprocal interactions}. We fix the parameters values $c_2 = 13$, $r = 1$, $s = 1$ and $a = 0.05$, and we show the bifurcation region as a function of the remaining parameters $d$ and $c_1$. The C region (grey) corresponds to an unstable homogeneous equilibrium that also remain unstable once the coupling is present, being the latter reciprocal or non-reciprocal one; patterns can develop but they are not due to the interactions. For parameters values in the A region (black), the stable homogeneous equilibrium is destabilised by the introduction of reciprocal coupling (as well by a non-reciprocal one); the patterns shown in the panel b) of Fig.~\ref{fig:patcmplxsymm} associated to the values $c_1=2$ and $d=0.02$ fall in this class (yellow dot). Finally in the large B region (white) the homogeneous equilibrium is stable for any reciprocal coupling, and thus no pattern can develop in this case, however the use of non-reciprocal interactions can drive the instability and thus the emergence of patterns. The patterns shown in the panel b) of Fig.~\ref{fig:patcmplx} (main text) correspond to the values $c_1=2$ and $d=0.4$ clearly belonging to the B region (blue star).}
\label{fig:BifDiag}
\end{figure}

In Fig.~\ref{fig:BifDiag} we provide a more global view of the parameters range associated to bifurcation diagram showing the parameters values, $d$ and $c_1$, for which the instability emerges in the case of both reciprocal interactions and non-reciprocal ones (black A region) and in the case of only non-reciprocal ones (white B region), once the remaining parameters have been fixed to some generic values. One can clearly appreciate how large is the latter compared to the former, and thus how more often one can find patterns due to non-reciprocal interactions instead of reciprocal ones.

\section{Analysis of the FitzHugh-Nagumo model}
\label{sec:FHNmod}

Let us consider the FitzHugh-Nagumo model, $\mathbf{f}(\mathbf{x}^{(i)})=(\mu u_i-u_i^3-v_i,\gamma(u_i-v_i))$, with $\mathbf{x}^{(i)}=(u_i,v_i)$, and the non-local coupling
\begin{equation*}
 \mathbf{F}(\mathbf{x}^{(i)},\mathbf{x}^{(j)})=(\mu u_i-u_i^3-v_j,\gamma(u_j-v_i))\, ,
\end{equation*}
with $\mathbf{x}^{(j)}=(u_j,v_j)$. From Eq.~\eqref{eq:sysnA} we obtain
\begin{equation*}
 \begin{dcases}
\frac{d u_i}{dt} &= \frac{1}{k_i^{(in)}}\sum_j A_{ij} \left(\mu u_i-u_i^3-v_j\right)\\
\frac{d v_i}{dt} &= \frac{1}{k_i^{(in)}}\sum_j A_{ij}\gamma\left( u_j-v_i\right)
\end{dcases} \quad\forall i=1,\dots, n\, .
\end{equation*}
By using again the definition of $k_i^{(in)}=\sum_j A_{ij}$ we get
\begin{equation*}
 \begin{dcases}
\frac{d u_i}{dt} &= \mu u_i-u_i^3-\frac{1}{k_i^{(in)}}\sum_j A_{ij} v_j\\
\frac{d v_i}{dt} &= -\gamma v_i +\frac{\gamma}{k_i^{(in)}}\sum_j A_{ij} u_j
\end{dcases} \quad\forall i=1,\dots, n\, ,
\end{equation*}
hence
\begin{equation*}
 \begin{dcases}
\frac{d u_i}{dt} &= \mu u_i-u_i^3-v_i+\sum_j \mathcal{L}_{ij}v_j\\
\frac{d v_i}{dt} &= \gamma (u_i-v_i) +\gamma \sum_j \mathcal{L}_{ij} u_j
\end{dcases} \quad\forall i=1,\dots, n\, .
\end{equation*}

The Jacobian matrices can be obtained as follows :
\begin{equation*}
 \mathbf{J}_{\mathrm{FHN}}=\left(
\begin{matrix}
\mu & -1\\ \gamma  &-\gamma
\end{matrix}
\right)\text{ and } \mathbf{J}_2=\left(
\begin{matrix}
0 & 1\\ \gamma &0
\end{matrix}
\right)\, .
\end{equation*}
By using the formulas~\eqref{eq:cS2} and~\eqref{eq:cS1}, we obtain the polynomials $S_1$ and $S_2$:
\begin{equation*}
S_1(\xi) =\gamma (\mu-\gamma)^2 \left[(1+\xi)^2-\mu\right]\text{ and }  S_2(\xi) = -4\gamma^2(\xi+1)^2-\gamma(\mu-\gamma)^2 \, .
\end{equation*}
The instability is thus realised if the complex eigenvalues $\Lambda^{(\alpha)}$ of $\mathcal{L}$ do satisfy the constraint
\begin{equation*}
(\Im \Lambda^{(\alpha)})^2 \geq \frac{(\mu-\gamma)^2}{4\gamma} \frac{(1+\Re \Lambda^{(\alpha)})^2-\mu}{(1+\Re \Lambda^{(\alpha)})^2+1}\, .
\end{equation*}

\section{Analysis of the Stuart-Landau model}
\label{sec:SLmod}

We turn now our attention to the study of the paradigmatic model of nonlinear oscillators given by the Stuart-Landau system (SL)~\cite{Stuart1978,Kuramoto}
\begin{equation*}
\frac{dw}{dt}=\sigma w-\beta w |w|^2\, ,
\end{equation*}
where $\sigma=\sigma_\Re+i\sigma_\Im$ and $\beta=\beta_\Re+i\beta_\Im$ are complex model parameters. One can straightforwardly prove that $w_{LC}(t)=\sqrt{\sigma_\Re/\beta_\Re}e^{i \omega t}$, $\omega=\sigma_\Im-\beta_\Im \sigma_\Re/\beta_\Re$, is a limit cycle solution of the SL system and it is stable provided $\sigma_\Re>0$ and $\beta_\Re>0$.

We then assume to have $n$ identical copies of the SL system coupled through non-reciprocal interactions, $A_{i\ell}\neq A_{\ell i}$, and the hypothesis of mean field Eq.~\eqref{eq:SL} (main text), hereby reported
\begin{equation*}
\frac{dw_j}{dt}=\frac{\sigma}{k^{(in)}_j}\sum_{\ell} A_{j\ell} w_\ell-\beta w_j|w_j|^2= \sigma w_j-\beta w_j|w_j|^2+\sigma \sum_{\ell} \mathcal{L}_{j\ell} w_\ell \, .
\end{equation*}

Because of the structure of the coupling, $w_{LC}(t)$ is also a solution of the latter equation. To inquire about its stability we consider the perturbation given by $w_j(t)=w_{LC}(t) \left(1+u_j(t)\right)e^{iv_j(t)}$, where $u_j(t)$ and $v_j(t)$ are real and small functions nodes dependent. We then insert the latter into~\eqref{eq:SL} (main text) and we expand by retaining only the first order terms by obtaining
\begin{eqnarray}
\label{eq:SLnospaceAlin}
\frac{d}{dt}\binom{u_j}{v_j}&=&\left(
\begin{matrix}
 -2\sigma_\Re & 0\\
 -2\beta_\Im \frac{\sigma_\Re}{\beta_\Re} & 0
\end{matrix}
\right)\binom{u_j}{v_j}+\sum_\ell \mathcal{L}_{j\ell}\left(
\begin{matrix}
\sigma_\Re & -\sigma_\Im\\
\sigma_\Im & \sigma_\Re
\end{matrix}
\right)\binom{u_\ell}{v_\ell}\, .
\end{eqnarray}

We invoke once again the existence of an eigenbasis of the Laplace matrix, $\phi^{(\alpha)}$, $\Lambda^{(\alpha)}$, to decompose the perturbation $u_j$ and $v_j$, and eventually get
\begin{eqnarray}
\label{eq:SLnospaceA2lin}
\frac{d}{dt}\binom{u^\alpha}{v^\alpha}&=&\left[\left(
\begin{matrix}
 -2\sigma_\Re & 0\\
 -2\beta_\Im \frac{\sigma_\Re}{\beta_\Re} & 0
\end{matrix}
\right)+\Lambda^{(\alpha)}\left(
\begin{matrix}
\sigma_\Re & -\sigma_\Im\\
\sigma_\Im & \sigma_\Re
\end{matrix}
\right)\right]\binom{u^\alpha}{v^\alpha}\notag\\
&=& \left(\mathbf{J}+\Lambda^{(\alpha)} \mathbf{J}_2\right)\binom{u^\alpha}{v^\alpha} =:\mathbf{J}^{(\alpha)} \binom{u^\alpha}{v^\alpha}\, .
\end{eqnarray}

By inserting the given expressions for $\mathbf{J}$ and $\mathbf{J}_2$, in the general formulas we obtain for the coefficients of $S_2(\xi)$
\begin{eqnarray}
 c_{2,2}&=& -\sigma_\Im^2 \left(\sigma_\Re^2 + \sigma_\Im^2\right) \notag\\
 c_{2,1}&=& 2 \sigma_\Im^2 \sigma_\Re \left(\beta_\Re \sigma_\Re + \beta_\Im \sigma_\Im\right)\frac{1}{\beta_\Re} \notag\\
 c_{2,0}&=& -\sigma_\Im^2 \sigma_\Re^2 \left(\beta_\Re^2 + \beta_\Im^2\right)\frac{1}{\beta_\Re^2} \, ,
 \end{eqnarray}
while for $S_1(\xi)$
\begin{eqnarray}
  c_{1,4}&=& \sigma_\Re^2 \left(\sigma_\Re^2 + \sigma_\Im^2\right) \notag\\
 c_{1,3}&=&-2 \sigma_\Re^2 \left(2 \beta_\Re \sigma_\Re^2 + \beta_\Im \sigma_\Im \sigma_\Re + \beta_\Re \sigma_\Im^2\right)\frac{1}{\beta_\Re} \notag\\
 c_{1,2}&=&\sigma_\Re^2 \left(5 \beta_\Re \sigma_\Re^2 + 4 \beta_\Im \sigma_\Im \sigma_\Re + \beta_\Re \sigma_\Im^2\right)\frac{1}{\beta_\Re} \notag\\
 c_{1,1}&=&-2 \sigma_\Re^3 \left(\beta_\Re \sigma_\Re + \beta_\Im \sigma_\Im\right)\frac{1}{\beta_\Re}\notag \\
 c_{1,0}&=&0 \, ,
 \end{eqnarray}

\begin{figure}[t]
\includegraphics[scale=.22]{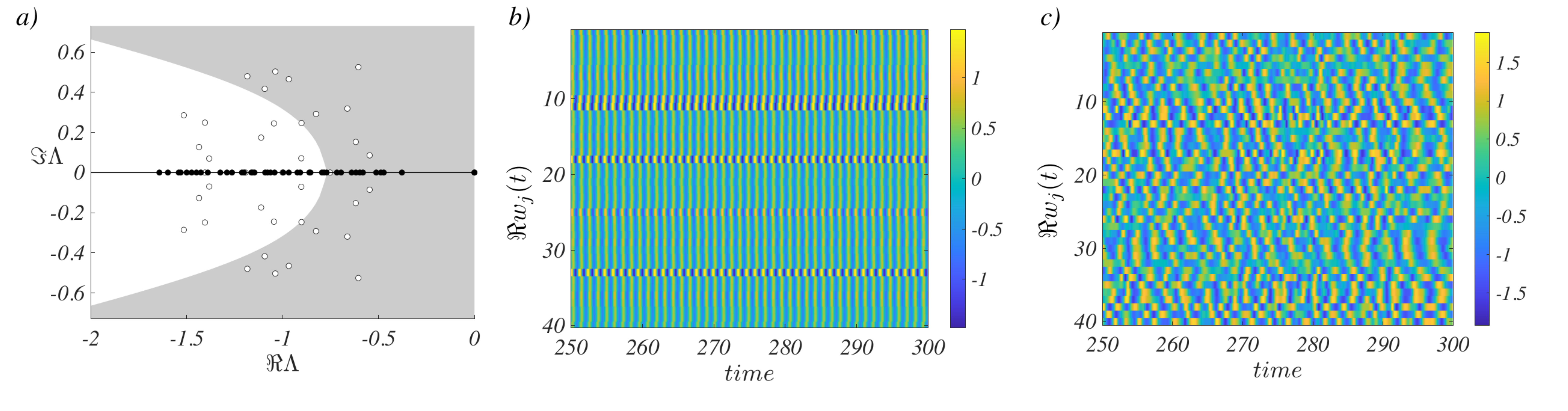}
\caption{\textbf{Instability region and patterns for the Stuart-Landau model}. In panel a) we report the region of the complex plane $(\Re \Lambda,\Im \Lambda)$ for which the instability condition is satisfied (grey), for the parameters values $\sigma_\Re = 1$, $\sigma_\Im=4.3$, $\beta_\Re=1$ and $\beta_\Im=-2$. We can observe that the instability region intersects the real axis and thus any kind of coupling, being reciprocal (black dots) or non-reciprocal one (white dots), can exhibit eigenvalues entering the instability region and thus allowing for an instability to set on, followed by a spatio-temporal patterns as shown in panel b) where we report the real part of the complex state variable $w_i$ in the case of a reciprocal coupling and panel c) for a non-reciprocal one).}
\label{fig:patcmplxSLsymm}
\end{figure}

As previously done in the case of the Volterra model, the explicit knowledge of the polynomial $S_1$ and $S_2$ allows to compute the (in)-stability region as shown in Fig.~\ref{fig:patcmplxSL} (main text) or Fig.~\ref{fig:patcmplxSLsymm} in a setting where the instability condition can be realised for both a reciprocal and non-reciprocal coupling. The instability region (grey) is shown in the complex plane $(\Re\Lambda,\Im\Lambda)$ together with the spectrum of a reciprocal web of long-range interactions (black dots) as well with a non-reciprocal one (white dots); because in both cases there are eigenvalues belonging to the instability region, the instability is possible and thus a spatio-temporal pattern emerges (see panel b) in the case of reciprocal can and panel c) for the non-reciprocal one). The numerical simulations have been perfumed using a $4$-th order Runge-Kutta method starting from initial conditions $\delta$-close to the homogeneous limit cycle solution $w_{LC}(t)=\sqrt{\sigma_\Re/\beta\Re}e^{i\omega t}$. In both cases the maximum of the dispersion relation $\rho_\alpha$ is of order of the unity and thus a relatively small integration span is sufficient to reveal the wavy solution. The underlying coupling is a directed Erd\H{o}s-R\'enyi network with $n=40$ nodes and a probability for a direct link to exist between two nodes is $p=0.08$.

\begin{figure}[t]
\includegraphics[scale=.30]{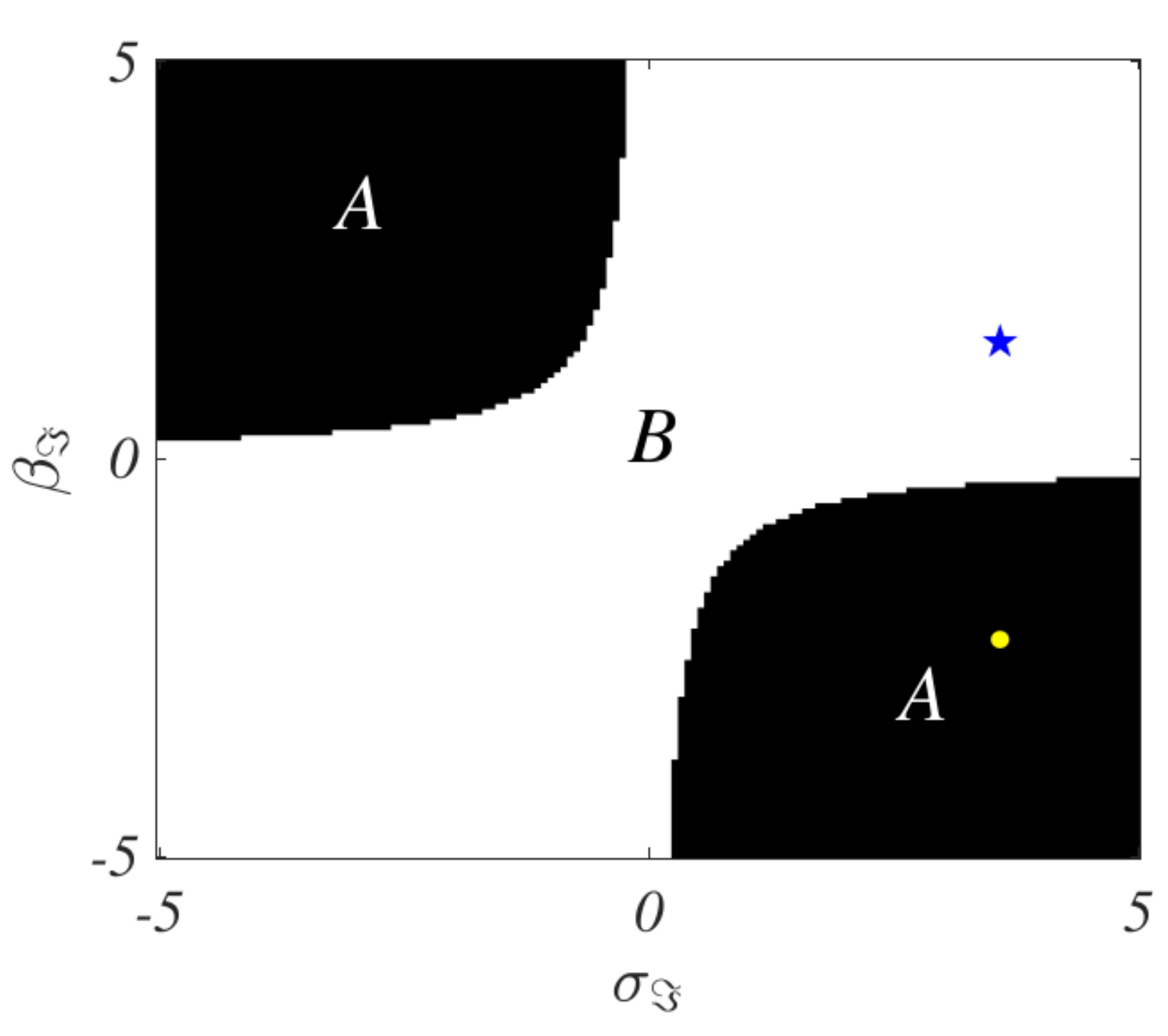}
\caption{\textbf{Bifurcation diagram for the Stuart-Landau model}. For the set of parameters values $\sigma_\Re = 1.0$ and $\beta_\Re = 1.0$ we report the range of the remaining parameters $\sigma_\Im$ and $\beta_\Im$ for which the instability can emerge. In the region A (black) the homogeneous equilibrium is stable and it can be destabilised by a reciprocal coupling as well by a non-reciprocal one. On the other hand, in the B region (white), the homogeneous equilibrium is always stable, no pattern can develop using a symmetric coupling. On the contrary one can found non-reciprocal coupling capable to destabilise the homogeneous equilibrium and thus the system to develop a wavy heterogeneous solution. The patterns shown in the panel b) of Fig.~\ref{fig:patcmplxSL} (main text) correspond to the values $\sigma_\Im = 4$ and $\beta_\Im = 1$ and clearly belong to the B region (blue star). The patterns presented in Fig.~\ref{fig:patcmplxSLsymm} for parameters values $\sigma_\Im = 4$ and $\beta_\Im = -2$ correspond to the region A (yellow dot).}
\label{fig:BifDiagSL}
\end{figure}

In Fig.~\ref{fig:BifDiagSL} we report the bifurcation diagram in the plane $\sigma_\Im$ and $\beta_\Im$, for $\sigma_\Re=\beta_\Re=1$. Two regions can be observed; in region A (black) the instability can be initiated by both a reciprocal and non-reciprocal web of long-range interactions while in region B (white) only non-reciprocal interactions can determine an instability and the ensuing wavy solution. 

\section{Instability onset in the general $d$-dimensional case}
\label{sec:generaldcase}

 {Let us consider $n$ copies of a generic $d$-dimensional system coupled through non-reciprocal interactions as in Eq.~\eqref{eq:sysnAL}, hereby recalled
\begin{equation*}
 \frac{d\mathbf{x}^{(i)}}{dt}=\mathbf{f}(\mathbf{x}^{(i)})+\sum_j \mathcal{L}_{ij} \mathbf{F}(\mathbf{x}^{(i)},\mathbf{x}^{(j)})\quad\forall i=1,\dots, n\, .
\end{equation*}
Assume there exists an homogeneous solution $\mathbf{x}^{(i)}=\mathbf{s}$ for the decoupled system, that clearly results also a solution for the coupled one. To study its stability in the coupled case, let us linearise the system about this reference solution to obtain~\eqref{eq:linear}, that can be further analysed by projecting it onto the Laplace eigenbasis to eventually get~\eqref{eq:finalEqalphaHom}.}

 {The stability of the reference solution can be determined by studying the sign of the real part of the roots of the characteristic polynomial~\eqref{eq:dispreld}
\begin{equation*}
p_d(\lambda):=(-1)^d\left[ a_0 \lambda^d+a_1\lambda^{d-1}+a_2\lambda^{d-2}+\dots+a_d\right]\, .
\end{equation*}
A straightforward computation, ultimately based on the expansion of the determinant defining the characteristic polynomial, allows to determine
\begin{eqnarray}
a_1 &=& -\mathrm{tr} \left(\mathbf{J}_1+\Re\Lambda^{(\alpha)}\mathbf{J}_2\right)-i \Im\Lambda^{(\alpha)} \mathrm{tr} \mathbf{J}_2\label{eq:a1}\\ 
 a_2 &=& \sum_{\ell =1}^d \sum_{j=\ell+1}^d \Big\{ (\mathbf{J}_1+\Re\Lambda^{(\alpha)}\mathbf{J}_2)_{\ell \ell}(\mathbf{J}_1+\Re\Lambda^{(\alpha)}\mathbf{J}_2)_{j j}-(\mathbf{J}_1+\Re\Lambda^{(\alpha)}\mathbf{J}_2)_{\ell j}(\mathbf{J}_1+\Re\Lambda^{(\alpha)}\mathbf{J}_2)_{j \ell}+\notag
 \\&+&(\Im \Lambda^{(\alpha)})^2\left[(\mathbf{J}_2)_{j \ell}(\mathbf{J}_2)_{\ell j}-(\mathbf{J}_2)_{\ell\ell}(\mathbf{J}_2)_{j j}\right]\Big\}+\notag\\
 &+&i\Im \Lambda^{(\alpha)} \sum_{\ell =1}^d \sum_{j=\ell+1}^d 
 \Big\{ (\mathbf{J}_1+\Re\Lambda^{(\alpha)}\mathbf{J}_2)_{\ell \ell}(\mathbf{J}_2)_{j j}+ (\mathbf{J}_1+\Re\Lambda^{(\alpha)}\mathbf{J}_2)_{j j}(\mathbf{J}_2)_{\ell \ell}- (\mathbf{J}_1+\Re\Lambda^{(\alpha)}\mathbf{J}_2)_{\ell j}(\mathbf{J}_2)_{j \ell}+\notag
 \\
 &-& (\mathbf{J}_1+\Re\Lambda^{(\alpha)}\mathbf{J}_2)_{j \ell}(\mathbf{J}_2)_{\ell j}\Big\}\label{eq:a2}\, .
\end{eqnarray}
Assuming to deal with reciprocal interactions, we can show (see Appendix~\ref{sec:spectLapsymm}) that the spectrum is real even  if the Laplace matrix is not symmetric. The previous equations simplify by imposing $\Im \Lambda^{(\alpha)}=0$, and in particular $a_1$ and $a_2$ are real numbers, that are assumed to be positive because of the hypothesis of stable equilibrium under symmetric interactions.}

 {Let us now discuss the role of the auxiliary polynomial $q(\lambda)$. Let $P(z)$ by a generic polynomial with complex coefficients. Assume $z_1$ to be a simple complex root and to simplify let us assume to be able to factorise $P(z)=(z-z_1)P_1(z)$ where $P_1(z)$ has real coefficients. Let us define $Q(z)=(z-\overline{z_1})P(z)$, then
\begin{equation*}
 Q(z)=(z-\overline{z_1})P(z)=(z-\overline{z_1})(z-z_1)P_1(z)=(z^2-2z \Re z_1+|z_1|^2)P_1(z)\, ,
\end{equation*}
from which it follows that $Q(z)$ has real coefficients and its roots have the same real part of the roots of $P(z)$ (computed twice in the case of $z_1$). So the general strategy would be to factorise all the complex roots of $P(z)$ and build an auxiliary polynomial $Q(z)$ according to this recipe. In general we do not know all such roots and thus we can simply assume to apply this process to all roots, even for the real ones, hence to the whole polynomial, and thus to define $Q(z)=\overline{P(\overline{z})}P(z)$. The resulting polynomial will have a degree equals to the double of the degree of $P(z)$, all its coefficients will be real by construction and its roots will have the same real part of the roots of $P(z)$.}

 {By applying such recipe to $p_{d}(\lambda)=\lambda^{d}+a_1\lambda^{d-1}+a_2\lambda^{d-2}+\dots+a_{d}$ we obtain the polynomial $q(\lambda)=\lambda^{2d}+b_1\lambda^{2d-1}+b_2\lambda^{2d-2}+\dots+b_{2d}$ defined in the main text. A direct computation consisting in equating coefficients of equal powers of $\lambda$ in $q(z)$ and $\overline{p_d(\overline{z})}p_d(z)$ allows to determine
\begin{eqnarray*}
 b_1&=& a_1+\overline{a_1}=2\Re a_1\, ,\\
 b_2&=&a_2+\overline{a_2}+a_1\overline{a_1}=2\Re a_2+|a_1|^2\, .
\end{eqnarray*}
By using the previously obtained expression~\eqref{eq:a1} for $a_1$, we can get
\begin{equation*}
 b_1=-2\mathrm{tr}\left(\mathbf{J}_1+\Re\Lambda^{(\alpha)}\mathbf{J}_2\right)\, ,
\end{equation*}
and we can observe that it coincides (module a factor $2$) with $a_1$ obtained once we impose $\Im \Lambda^{(\alpha)}=0$; hence $b_1$ results to be always positive, as $a_1$ does. The non-reciprocal interactions cannot thus change the sign of $b_1$. On the other hand by using the expression for $a_2$ given in~\eqref{eq:a2}, we can obtain the conditions presented in the main text~\eqref{eq:consign} and~\eqref{eq:condx} ensuring $b_2<0$. Indeed recalling~\eqref{eq:a2} we can write
\begin{equation*}
b_2 =  2\Re a_2+|a_1|^2 = \left(2\Re a_2+|a_1|^2\right)_{\Im \Lambda^{(\alpha)}=0}+2(\Im \Lambda^{(\alpha)})^2\sum_{\ell =1}^d \sum_{j=\ell+1}^d\Big\{\left[(\mathbf{J}_2)_{j \ell}(\mathbf{J}_2)_{\ell j}-(\mathbf{J}_2)_{\ell\ell}(\mathbf{J}_2)_{j j}\right]\Big\}+(\Im \Lambda^{(\alpha)})^2 \left(\mathrm{tr}\mathbf{J}_2\right)^2\, .
\end{equation*}
By observing that $\left(2\Re a_2+|a_1|^2\right)_{\Im \Lambda^{(\alpha)}=0}>0$ because of the assumption of stability of the system using symmetric interaction, we can conclude that $b_2$ can be negative if $(\Im \Lambda^{(\alpha)})^2$ is large enough and
\begin{equation*}
2\sum_{\ell =1}^d \sum_{j=\ell+1}^d\Big\{\left[(\mathbf{J}_2)_{j \ell}(\mathbf{J}_2)_{\ell j}-(\mathbf{J}_2)_{\ell\ell}(\mathbf{J}_2)_{j j}\right]\Big\}+\left(\mathrm{tr}\mathbf{J}_2\right)^2<0\, .
\end{equation*}
After some algebraic manipulations, we can rewrite the term involving the double sum as
\begin{equation*}
\sum_{\ell =1}^d \sum_{j=\ell+1}^d\Big\{\left[(\mathbf{J}_2)_{j \ell}(\mathbf{J}_2)_{\ell j}-(\mathbf{J}_2)_{\ell\ell}(\mathbf{J}_2)_{j j}\right]\Big\} = \frac{\mathrm{tr}\mathbf{J}^2_2 - \left(\mathrm{tr}\mathbf{J}_2\right)^2}{2}\, ,
\end{equation*}
from which the condition~\eqref{eq:consign} follows. This is a sufficient conditions for the destabilisation of the equilibrium solution; if the latter does not hold true one could in principle look for other conditions capable to change the sign of one of the remaining coefficients, $b_j$, of the polynomial $q(\lambda)$.}

\end{document}